\documentclass{article}\usepackage[utf8]{inputenc}
\pdfoutput=1
\usepackage{amsmath, amssymb, amsthm, amscd, amsfonts}
\usepackage{url}
\usepackage{xcolor}
\usepackage{graphicx}
\usepackage[margin=1in]{geometry}

\usepackage{verbatim}
\usepackage{algorithm}
\usepackage[noend]{algpseudocode}
\usepackage{float}
\usepackage{multicol}
\usepackage{xspace}
\usepackage{wrapfig}

\usepackage{amsfonts}
\usepackage{graphicx}
\usepackage{subcaption}
\usepackage{url}
\usepackage{times}
\usepackage{enumerate}
\usepackage{venndiagram}
\usepackage{import}
\usepackage{xcolor}

\usepackage{thmtools,thm-restate}

\theoremstyle{plain} \numberwithin{equation}{section}
\newtheorem{theorem}{Theorem}[section]

\newtheorem{corollary}[theorem]{Corollary}

\newtheorem{lemma}[theorem]{Lemma}

\theoremstyle{definition}
\newtheorem{definition}[theorem]{Definition}

\newcommand{\cmdclass}[0]{\textbf{class}\xspace}

\newcommand{\cmdconstructor}[0]{\textbf{constructor}\xspace}
\newcommand{\cmdprocedure}[0]{\textbf{procedure}\xspace}
\newcommand{\cmdstatic}[0]{\textbf{static}\xspace}

\newcommand{\cmdif}[0]{\textbf{if}\xspace}
\newcommand{\cmdthen}[0]{\textbf{then}\xspace}

\newcommand{\cmdelse}[0]{\textbf{else}\xspace}

\newcommand{\cmdreturn}[0]{\textbf{return}\xspace}

\newcommand{\cmdfor}[0]{\textbf{for}\xspace}

\newcommand{\cmdnew}[0]{\textbf{new}\xspace}
\newcommand{\cmdstruct}[0]{\textbf{struct}\xspace}
\newcommand{\cmdinstancevar}[0]{\textbf{instance variable}\xspace}

\newcounter{int}
\newcommand{\tab}[1][1]{\setcounter{int}{0}\loop\hspace{\algorithmicindent}\addtocounter{int}{1}\ifnum\value{int}<#1\repeat}

\newcommand{\true}[0]{\mbox{\em true}\xspace} 
\newcommand{\false}[0]{\mbox{\em false}\xspace}

\newcommand{\CAS}[1]{\Call{Cas}{#1}}

\newcommand{\CASObj}[1]{{#1}}
\newcommand{\X}[0]{\CASObj{X}}
\newcommand{\Y}[0]{\CASObj{Y}}

\newcommand{\DataObj}[1]{\mathcal{#1}}
\renewcommand{\O}[0]{\DataObj{O}}

\newcommand{\Z}[0]{\DataObj{Z}}
\newcommand{\W}[0]{\DataObj{W}}

\newcommand{\SharedVar}[1]{\mbox{\text{#1}}}
\newcommand{\StructField}[1]{\mbox{\textit{#1}}}

\newcommand{\Val}[0]{\SharedVar{\StructField{Val}}}
\newcommand{\Last}[0]{\SharedVar{\StructField{DetVal}}}

\newcommand{\Critical}[0]{\SharedVar{\StructField{Critical}}}
\newcommand{\Casual}[0]{\SharedVar{\StructField{Casual}}}
\newcommand{\ECWH}[0]{\SharedVar{\StructField{ECWH}}}

\newcommand{\pc}{pc}

\newcommand  {\h}[1][]{h_{#1}}
\newcommand  {\s}[1][]{s_{#1}}
\renewcommand{\v}[1][]{v_{#1}}
\newcommand  {\w}[1][]{w_{#1}}
\newcommand  {\x}[1][]{x_{#1}}
\newcommand  {\y}[1][]{y_{#1}}
\newcommand  {\z}[1][]{z_{#1}}

\newcommand  {\sh}[1][]{\Hat{\s}_{#1}}
\newcommand  {\hh}[1][]{\Hat{\h}_{#1}}
\newcommand  {\vh}[1][]{\Hat{\v}_{#1}}
\newcommand  {\wh}[1][]{\Hat{\w}_{#1}}
\newcommand  {\zh}[1][]{\Hat{\z}_{#1}}

\newcommand{\old}[1][]{old_{#1}}
\newcommand{\new}[1][]{new_{#1}}
\newcommand{\ret}[1][]{r_{#1}}
\newcommand{\initval}[1][]{initval_{#1}}

\newcommand{\val}[0]{val}
\newcommand{\seq}[0]{seq}
\newcommand{\hndl}[0]{hndl}
\newcommand{\bit}[0]{bit}

\newcommand{\context}[0]{context}
\newcommand{\contexts}[0]{contexts}
\newcommand{\self}[0]{\mbox{\em self}}

\newcommand{\detval}[0]{\Last}

\newcommand{\cas}[0]{CAS}

\newcommand{\Type}[1]{\textbf{#1}}

\newcommand{\Tint}{\Type{int}}
\newcommand{\Thandle}{\Type{handle*}}

\newcommand{\Obj}[1]{\mbox{\textsc{#1}}}    
\newcommand{\Method}[1]{\mbox{\textsc{#1}}} 
\newcommand{\method}[1]{\mbox{\textsc{#1}}} 

\newcommand{\Recover}[2][]{\Method{Recover}_{#1}(#2)}
\newcommand{\Detect}[2][]{\Method{Detect}_{#1}(#2)}

\newcommand{\CreateHandle}[1]{\Method{CreateHandle}(#1)}

\newcommand{\Read}[1]{\Method{Read}(#1)}
\newcommand{\Write}[1]{\Method{Write}(#1)}

\newcommand{\ECL}[1]{\textsc{DurEC}(#1)}
    \newcommand{\DurEC}[0]{\Obj{DurEC}}
        \newcommand{\CLL}[1]{\Method{ECLL}(#1)}
        \newcommand{\CVL}[1]{\Method{ECVL}(#1)}
        \newcommand{\CSC}[1]{\Method{ECSC}(#1)}

    \newcommand{\DuraCAS}[0]{\Obj{DuraCAS}}   
        \newcommand{\Rcas}[1]{\Method{CAS}(#1)}

            \newcommand{\transfer}[2][]{\method{transfer-write}_{#1}(#2)}
            \newcommand{\forward}[2][]{\method{forward}_{#1}(#2)}

    \newcommand{\DuraLL}[0]{\Obj{DuraLL}}
        \newcommand{\LL}[1]{\Method{LL}(#1)}
        \newcommand{\VL}[1]{\Method{VL}(#1)}
        \newcommand{\SC}[1]{\Method{SC}(#1)}

    \newcommand{\DurECW}[0]{\Obj{DurECW}}

        \newcommand{\Remove}[1]{\Method{Remove}(#1)}

\title{Durable Algorithms for Writable LL/SC and CAS with Dynamic Joining}

\author{
Prasad Jayanti\footnote{Dartmouth College; prasad.jayanti@dartmouth.edu} 
\and 
Siddhartha V. Jayanti\footnote{Google Research and MIT; sjayanti@google.com, siddhartha@csail.mit.edu} 
\and 
Sucharita L. Jayanti\footnote{Brown University; sucharita\_jayanti@brown.edu} 
}

\date{January 10, 2023}

\begin{document}

\maketitle
\begin{abstract}


We present durable implementations for two well known universal primitives---CAS (compare-and-swap), and its ABA-free counter-part LLSC (load-linked, store-conditional).
All our implementations are: {\em writable}, meaning they support a Write() operation; have {\em constant time complexity} per operation; allow for {\em dynamic joining}, meaning newly created processes (a.k.a. threads) of arbitrary names can join a protocol and access our implementations; and have {\em adaptive space complexities}, meaning the space use scales in the number of processes $n$ that actually use the objects, as opposed to previous protocols which are designed for a maximum number of processes $N$.
Our durable Writable-CAS implementation, DuraCAS, requires $O(m + n)$ space to support $m$ objects that get accessed by $n$ processes, improving on the state-of-the-art $O(m + N^2)$.
By definition, LLSC objects must store ``contexts'' in addition to object values.
Our Writable-LLSC implementation, DuraLL, requires $O(m + n + C)$ space, where $C$ is the number of ``contexts'' stored across all the objects.
While LLSC has an advantage over CAS due to being ABA-free, the object definition seems to require additional space usage.
To address this trade-off, we define an {\em External Context (EC)} variant of LLSC.
Our EC Writable-LLSC implementation is ABA-free and has a space complexity of just $O(m + n)$.

To our knowledge, we are the first to present durable CAS algorithms that allow for dynamic joining, and our algorithms are the first to exhibit adaptive space complexities.
To our knowledge, we are the first to implement any type of {\em durable} LLSC objects.

\end{abstract}

\section{Introduction}
\label{sec:intro}

The advent of {\em Non-Volatile Memory (NVM)} \cite{Optane} has spurred the development of durable algorithms for the {\em crash-restart model}.
In this model, when a process $\pi$ crashes,
the contents of memory {\em persist} (i.e., remain unchanged), but $\pi$'s cache and CPU registers lose their contents, and its program counter is set to a default value upon restart.
To understand the difficulty that arises from losing register contents,
suppose that $\pi$ crashes at the point of executing a hardware CAS instruction, $r \gets \CAS{X, \old, \new}$,
on a memory word $X$ and receiving the response into its CPU register $r$.
When $\pi$ subsequently restarts,
$\pi$ cannot tell whether the crash occurred before or after the CAS executed, and if the crash occurred after the CAS, $\pi$ cannot tell whether the CAS was successful or not.
Researchers identified this issue and proposed software-implemented {\em durable objects} \cite{Izraelevitz, Attiya}, which allow a restarted process to {\em recover} from its crash and {\em detect} the result of its last operation.
This is done by exposing two additional methods, $\Recover{}$ and $\Detect{}$.
The rapid commercial viability of byte-addressable, dense, fast, and cheap NVM chips has made efficient durable object design important.

\paragraph{\bf Writable and non-Writable CAS}
Recently, there has been a lot of research on implementing durable CAS objects because they are widely employed in practice and are universal; any durable object can be implemented from durable CAS objects \cite{wait-free, Izraelevitz, BenDavid}.
Formally, the state of a CAS object $X$ is simply its value, and the operation semantics are as follows:
\begin{itemize}
\vspace{-0.03in}
    \item 
    $X.\CAS{\old, \new}$: if $X = \old$, sets $X$ to $\new$ and returns $\true$; otherwise, returns $\false$.

    \item
    $X.\Read{}$: returns the value of $X$.

    \item
    $X.\Write{\new}$: sets $X$ to $\new$ and returns $\true$.
\end{itemize}
\vspace{-0.03in}
If the object supports all three operations, it is a {\em Writable-CAS (W-CAS)}, if it does not support $\Write$, it is a {\em non-Writable-CAS (nW-CAS)} object.

\paragraph{\bf CAS's ABA problem and LLSC}
Although CAS objects are powerful tools in concurrent computing, they also have a significant drawback called the {\em ABA-problem} \cite{Bjarne}.
Namely, if a process $\pi$ reads a value $A$ in $X$ and executes $X.\CAS{A, C}$ at a later time, this CAS will succeed {\em even if} the value of $X$ changed between $\pi$'s operations, from $A$ to $B$ and then back to $A$.
So while any object \textit{can} be implemented from CAS, the actual process of designing an algorithm to do so becomes difficult.
In the non-durable setting, the ABA-problem is often overcome by using the hardware's double-width CAS primitive---in fact, `CAS2 [double-width CAS] operation is the most commonly cited approach for ABA prevention in the literature'' \cite{Bjarne}.
However, all known durable CAS objects, including ours, are only one-word wide---even as they use hardware double-width CAS \cite{Attiya, BenDavid, BenBaruch}.
Against this backdrop, the durable LLSC objects presented in this paper serve as an invaluable alternate tool for ABA prevention. 

LLSC objects are alternatives to CAS objects that have been invaluable in practice, since they are universal and ABA-free \cite{JayantiPetrovic2005}. 
The state of an LLSC object $Y$ is a pair $(Y.\val, Y.\context)$, where $Y.\val$ is the {\em value} and $Y.\context$ is a set of processes (initially empty).
Process $\pi$'s operations on the object have the following semantics:
\begin{itemize}
\vspace{-0.03in}
\item 
$Y.\LL{}$: adds $\pi$ to $X.\context$ and returns $Y.\val$.
\item 
$Y.\VL{}$: returns whether $\pi \in Y.\context$.
\item
$Y.\SC{\new}$: if $\pi \in X.\context$, sets $Y$'s value to $\new$, resets $Y.\context$ to the empty set and returns $\true$; otherwise, returns $\false$.
\item
$Y.\Write{\new}$ changes $Y$'s value to $\new$ and resets $Y.\context$ to the empty set.
\end{itemize}
\vspace{-0.03in}
The object is {\em Writable (W-LLSC)} or {\em non-Writable (nW-LLSC)} depending on whether the $\Write$ operation is supported.

To our knowledge, there are no earlier durable implementations of ABA-free CAS-like objects, including LLSC.

\paragraph{\bf Previous work and the state-of-the-art}
CAS and LLSC objects share close ties, but they also pose different implementational challenges.
In the non-durable context, it is well known that non-writable LLSC (nW-LLSC) objects can be implemented from nW-CAS objects and visa versa in constant time and space.
The simple implementation of nW-LLSC from nW-CAS however, requires packing a value-context pair into a single nW-CAS object \cite{LLSCfromCAS}.
Solutions that implement a full-word nW-LLSC from a full-word nW-CAS require a blow-up in time complexity, space complexity, or both \cite{JayantiPetrovic2003, DohertyHerlihyLuchangoMoir2004, Michael2004, JayantiPetrovic2005, BlellochWeiLLSCfromCAS}.
Writability complicates the relationship further. 
Even in the non-durable context, reductions between W-CAS and W-LLSC have resulted in a blow-up in space complexity and fixing the number of processes {\em a priori} \cite{WritableLLSCfromCAS}. 
Writability can sometimes be added to an object that is non-writable, but this leads to an increase in space complexity \cite{Aghazadeh}.

There are no previous works on Durable LLSC.
Three previous works have implemented durable CAS objects, all from the hardware CAS instruction: Attiya et al. \cite{Attiya}, Ben-Baruch et al. \cite{BenBaruch}, and Ben-David et al. \cite{BenDavid}.
All three papers provide implementations for a fixed set of $N$ processes with $pid$s $1,\ldots,N$, and achieve constant time complexity per operation.
Attiya et al. pioneered this line of research with a durable nW-CAS implementation, which achieves constant time complexity and requires $O(N^2)$ space per object.
Ben-Baruch et al. present an nW-CAS implementation with optimal bit complexity.
Their algorithm however, requires packing $N$ bits and the object's value into a single hardware variable.
Thus, if the value takes 64 bits, then only 64 pre-declared processes can access this object. 
(Current commodity multiprocessors range up to 224 cores \cite{224-core}, and can support orders-of-magnitude more threads.)
Ben-David et al. designed an algorithm for nW-CAS, and then leveraged Aghazadeh et al.'s writability transformation \cite{Aghazadeh} to enhance that algorithm to include a Write operation, thereby presenting the only previous Writable-CAS implementation.
Their nW-CAS algorithm uses a pre-allocated help-array of length $O(N)$, and their W-CAS algorithm uses an additional hazard-pointer array of length $O(N^2)$. 
Both arrays can be shared across objects, thus the implementation space complexities for $m$ objects are $O(m + N)$ and $O(m + N^2)$, respectively.
\paragraph{\bf Our contributions}
We present four wait-free, durably linearizable implementations: $\DuraCAS$ for Writable-CAS, $\DuraLL$ for Writable-LLSC, $\DurEC$ for External Context (EC) nW-LLSC, and $\DurECW$ for EC W-LLSC (the last two are described in the section below).
Our implementations achieve the following properties:
\begin{enumerate}       
    \item \underline{Constant time complexity}: 
    all operations including recovery and detection run in $O(1)$ steps.

    \item \underline{Dynamic Joining}:         
    dynamically created processes of arbitrary names can use our objects.

    \item \underline{Full-word size}:
    Our implementations support full-word (i.e., 64-bit) values.

    
    \item \underline{Adaptive Space Complexity}:
    We quantify space complexity by the number of memory words needed to support $m$ objects for a total of $n$ processes.
    The $\DuraCAS$, $\DurEC$, and $\DurECW$ implementations require just constant memory per process and per object, and thus each have a space complexity of $O(m + n)$.
    Since $\DuraLL$ must remember contexts, its space complexity is $O(m + n + C)$, where $C$ is the number of contexts that must be remembered\footnote{$C$ is the number of process-object pairs $(\pi, \O)$, where $\pi$ has performed an $\LL$ operation on $\O$, and its last operation on $\O$ is not an $\SC$ or $\Write$. A trivial upper bound is $C \le nm$.}.

    \end{enumerate}
We believe that our definitions and implementations of the External Context LLSC objects---which are ABA-free, space-efficient alternatives to CAS and LLSC---are of independent interest in the design of both durable and non-durable concurrent algorithms.

To our knowledge, we are the first to present durable CAS algorithms that allow for dynamic joining, and our algorithms are the first to exhibit adaptive space complexities.
To our knowledge, we are the first to consider any type of {\em durable} LLSC objects.

\paragraph{\bf Our approach}
We implement universal primitives that allow dynamic joining of new processes, have an {\em adaptive space complexity} that is constant per object and per process, and give an ABA-free option, while simultaneously achieving constant time complexity.
Just like our predecessors, all our implementations rely on just the hardware double-width CAS instruction for synchronization.

A keystone of our approach is the observation that durable nW-LLSC---due to its ABA-freedom---serves as a better stepping stone than even durable nW-CAS on the path from hardware CAS to durable W-CAS.
Perhaps less surprisingly, durable nW-LLSC is a great stepping stone towards durable W-LLSC also.
However, by definition LLSC objects require more space to remember context for each process---an inherent burden that CAS objects do not have.
Thus, using nW-LLSC objects in the construction of our W-CAS would lead to a bloated space complexity.
To avoid this drawback, we define an {\em External Context (EC)} variant of LLSC.
An EC LLSC object is like an LLSC object, except that its context is returned to the process instead of being maintained by the object.
Thus, our EC nW-LLSC implementation, $\DurEC$, is the building block of all our other implementations. 

The state of an EC LLSC object $Y$ is a pair $(Y.\val, Y.\seq)$, where the latter is a sequence number context.
Process $\pi$'s operations on the object have the following semantics:
\begin{itemize}
\vspace{-0.03in}
\item 
$Y.\CLL{}$: returns ($Y.\val$, $\Y.\seq)$.
\item 
$Y.\CVL{s}$: returns whether $\Y.\seq = s$.
\item
$Y.\CSC{s, \new}$: if $\Y.\seq = s$, sets $Y$'s value to $\new$, increases $\Y.\seq$, and returns $\true$; otherwise, returns $\false$.
\item
$Y.\Write{\new}$: changes $Y$'s value to $\new$ and increases $\Y.\seq$.
\end{itemize}
\vspace{-0.03in}
The object is {\em Writable (EC W-LLSC)} or {\em non-Writable (EC nW-LLSC)} depending on whether the $\Write$ operation is supported.

\begin{figure}[h]
\centering
\begin{subfigure}[b]{0.8\textwidth}
   \includegraphics[width=1\linewidth]{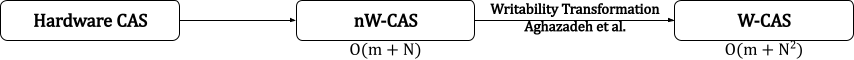}
\end{subfigure}

\begin{subfigure}[b]{0.8\textwidth}
   \includegraphics[width=1\linewidth]{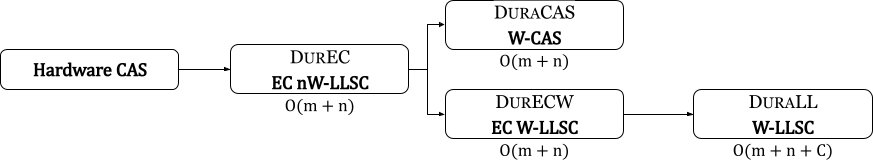}
\end{subfigure}

\caption[Two approaches]{
A comparison of Ben-David et al.'s approach (top) and our approach (bottom): 
each box represents an implementation---the type of the implementation is in bold and its space complexity appears below the box.
The names of our implementations appear in the box in {\sc SmallCaps}.
An arrow from A to B means that B is implemented using A.}
\label{fig:approaches}
\end{figure}

We design durable implementations of EC W-LLSC and W-CAS, called $\DurECW$ and $\DuraCAS$, respectively; each implementation uses two $\DurEC$ base objects.
We implement our durable W-LLSC algorithm, $\DuraLL$, by simply internalizing the external contexts of a $\DurECW$.
All our implementations overcome the need for hazard-pointers and pre-allocated arrays for helping in order to allow dynamic joining and achieve adaptive space complexity.
Key to eliminating these arrays are pointer based identity structures called {\em handles}, which we showcase in the next section.
Figure~\ref{fig:approaches} illustrates the differences between our approach and Ben-David et al.'s.

\subsection{Other Related Work}
\label{sec:related-work}


Byte-addressable non-volatile memory laid the foundation for durable objects \cite{Optane}.
Research on durable objects has spanned locks \cite{GolabRamaraju, ramaraju:rglock, JayantiJoshi1, JayantiJoshi2, JayantiJayantiJoshi1, JayantiJayantiJoshi2, GolabHendler, Bohannon:userlevel, Chan, Mittal, ChanWoelfelLB}, and non-blocking objects---including queues \cite{HerlihyQueue}, counters \cite{Attiya}, registers \cite{Attiya, BenBaruch}, and CAS objects \cite{Attiya, BenBaruch, BenDavid}.
The correctness criterion for non-blocking objects, {\em durable linearizability}, was first introduced for the full-system-crash model by Izraelevitz et al. \cite{Izraelevitz}, and adapted to the individual process crash-restart model used in this paper by Attiya et al. \cite{Attiya}.
Several other works have explored variants of the durable linearizability definition \cite{HerlihyQueue, StrictLinearizability, BDFW2022, DetectableSequentialSpecs, DBLP:conf/opodis/BerryhillGT15, BenDavid}.

\section{Model}

We use the {\em crash-restart model} with independent process crashes \cite{Izraelevitz, GolabRamaraju, Attiya, HerlihyQueue, BenDavid, BenBaruch}.
In this model, asynchronous processes communicate by applying atomic operations to Non-Volatile Memory (NVM).
Our algorithms use the read and compare-and-swap (CAS) operations.
Any process may crash at any time and restart at any later time, and the same process may crash and restart any number of times.
When a process crashes, its registers, including its program counter, lose their contents (i.e., they are set to arbitrary values), but the contents of the NVM are unaffected.

A  {\em durable implementation} of an object $\O$ provides one method for each operation supported by $\O$ and two additional methods---$\O.\Recover$ and $\Detect$.
If a process $p$ invokes a method for an operation and completes the method without crashing, the operation is required to take effect atomically at some instant between the method's invocation and completion.
On the other hand, if $p$ crashes while executing the operation, when $p$ subsequently restarts, it is required to execute $\O.\Recover$---if it crashes while executing the recover method, it must re-execute $\O.\Recover$ when it restarts.
The crashed operation is considered {\em complete} when $\O.\Recover$ completes.
The correctness condition for these methods is durable linearizability \cite{Izraelevitz, Attiya}, which generalizes linearizability \cite{linearizability} to the crash-restart model, and is stated as follows. 
The crashed operation is required to either have no effect at all or take effect atomically at some instant between when the method for the operation is invoked and when the recover method completes.

In addition to being durable, the objects implemented in this paper are also {\em detectable} \cite{HerlihyQueue}.
Detectability provides a means for processes to distinguish whether their crashed operations (that have subsequently been completed via the recover method) have taken effect or not, and what the associated response was.
Some operations, such as read or a failed CAS, can safely be repeated, regardless of whether they took effect \cite{Attiya, BenDavid}.
On the other hand, a write or a successful CAS that changed the value of the object cannot be repeated safely; such visible operations should be detected.
The $\Detect$ method facilitates detectability.
A call to the $\Detect$ method by a process $p$ returns a pair $(d, r)$, where $d$ is a {\em detection value} corresponding to the last detected operation by $p$ and $r$ is that operation's response.
Specifically, if $p$ calls $\Detect$ twice---just before executing an operation and just after completing that operation---and these successive calls to $\Detect$ return $(d_1, r_1)$ and $(d_2, r_2)$ respectively, then the following two properties are satisfied:
\begin{enumerate}
    \item 
    If $d_2 > d_1$, then the operation took effect and its response is $r_2$.    
    \item
    Otherwise, $d_2 = d_1$ and the operation is safe to repeat.
\end{enumerate}

\section{Handles for dynamic joining and space adaptivity}

When a process calls a method to execute an operation $op$, the call is of the form $op(p, args)$, where $args$ is a list of $op$'s arguments and $p$ identifies the calling process.
The methods use $p$ to facilitate helping between processes.
In many algorithms, the processes are given $pid$s from 1 to $N$, and $p$ is the $pid$ of the caller \cite{Attiya,BenDavid}.
In particular, $p$ is used to index a pre-allocated helping array---in Ben-David et al.'s algorithm this helping array is of length $N$, one location per process being helped; in Attiya et al.'s algorithm this helping array is of length $N^2$, one location per helper-helpee pair.
Helping plays a central role in detection, thus each process needs to have some area in memory where it can be helped; in fact, using the bit-complexity model, Ben-Baruch et al. proved that the space needed to support a detectable CAS object monotonically increases in the number of processes that access the object \cite{BenBaruch}.
One of our goals in this paper however, is to design objects that can be accessed by a dynamically increasing set of processes, which precludes the use of pre-allocated fixed-size arrays that are indexed by process IDs.

To eliminate the use of arrays for helping, we introduce pointer based structures called {\em handles}.
We use handles to enable dynamic joining and achieve space adaptivity. 
A handle is a constant sized record in memory.
The implementation provides a {\sc create-handle}$()$ method, which creates a new handle and returns a pointer $p$ to it.
When a process first wishes to access any of the implemented objects of a given type, it creates for itself a new handle by calling {\sc create-handle}$()$.
From that point on, whenever the process calls any method on any of the implemented objects of that type, it passes in the pointer $p$ of its handle instead of its pid, and other processes help it via the handle.
This mechanism of handles helps us realize dynamic joining because any number of new processes can join at any time by creating handles for themselves; since the memory per handle is constant, and only the subset of processes that wish to access the implementation need to create handles, the mechanism facilitates space adaptivity. 
\section{The $\DurEC$ Building Block}
\label{sec:DurEC}

In this section, we implement the $\DurEC$ algorithm for durable external context non-writable LLSC using hardware CAS.
This building block will be central to all of the writable implementations in the remainder of the paper.

\subsection{Intuitive description of Algorithm \DurEC}

Each DurEC handle $\h$ is a reference to a record of two fields, $\Val$ and $\detval$,
and each DurEC object $\O$ is implemented from two hardware atomic CAS objects $\X$ and $\Y$, where $\X$ is a pair consisting of a handle and a sequence number, and $\Y$ is a pair consisting of a sequence number and a value.
The algorithm maintains the DurEC object $\O$'s state in $\Y$, i.e., $\O.\seq = \Y.\seq$ and $\O.\val = \Y.\val$ at all times.
This representation makes the implementation of ECLL and ECVL operations obvious: $\CLL{\h}$ simply returns $\Y$ and $\CVL{\h, \s}$ returns whether $\Y.\seq = \s$.
The complexity lies in the $\CSC{\h, \s, \v}$ operation, which is implemented by the following sequence of steps:

\begin{enumerate}
\item
If $\Y.seq \neq \s$, it means $\O.\seq \neq \s$, so the ECSC operation simply returns $\false$. Otherwise, it embarks on the following steps, in an attempt to switch $\O.\val$ to $\v$ and $\O.\seq$ to a greater number.
\item
Make $v$ available for all by writing it in the $\Val$ field of the ECSC operation's handle $\h$.
\item
Pick a number $\sh$ that is bigger than both $\X.\seq$ and $\h.\detval$.
(The latter facilitates detection.)
\item
Publish the operation's handle along with a greater sequence number by
installing $(\h, \sh)$ in $\X$.
If several ECSC operations attempt to install concurrently, only one will succeed.
The successful one is the {\em installer} and the others are {\em hitchhikers}.
\item
The installer and the hitchhikers work together to accomplish two missions, the first of which is to increase the installer's $\detval$ field to the number in $\X.\seq$. This increase in the $\detval$ field of its handle enables the installer to detect that it installed, even if the installer happens to crash immediately after installing.
\item
The second mission is to forward the installer's operation to $\Y$.
Since $\Y$ is where the DurEC object's state is held, the installer's operation takes effect only when it is reflected in $\Y$'s state.
Towards this end, everyone reads the installer's value $\v$,  made available in the $\Val$ field of the installer's handle back at Step (2), and attempts to switch $\Y.\val$ to $\v$, simultaneously increasing $\Y.\seq$ so that it catches up with $\X.\seq$.
Since all operations attempt this update of $\Y$, someone (not necessarily the installer) will succeed.
At this point, $\X.\seq = \Y.\seq$ and $\Y.\val = \v$, which means that the installer's value $\v$ has made its way to $\O.\val$.
So, the point where $\Y$ is updated becomes the linearization point for the installer's successful ECSC operation.
The hitchhikers are linearized immediately after the installer, which causes their ECSC operations to ``fail''---return $\false$, without changing $\O$'s state---thereby eliminating the burden of detecting these operations.
\item
If the installer crashes after installing, upon restart, in the Recover method, it does the forwarding so that the two missions explained above are fulfilled.
\item
With the above scheme, all ECSC, ELL, and EVL operations, except those ECSC operations that install, are safe to return and hence, don't need detection.
Furthermore, for each installing ECSC operation, the above scheme ensures that the $\detval$ field of the installer's handle is increased, thereby making the operation detectable.
\end{enumerate}

The formal algorithm is presented in Figure~\ref{alg:DurEC}.
The correspondence between the lines of the algorithm and the steps above is as follows.
Lines 6 and 7 implement Steps 1 and 2, respectively.
Steps 3 and 4, where the operation attempts to become the installer, are implemented by Lines 8 to 10.
The operation becomes the installer if and only if the CAS at Line~10 succeeds, which is reflected in the boolean return value $r$.
The Forward method is called at Line~11 to accomplish the two missions described above.
The first three lines of Forward (Lines 13 to 15) implement the first mission of increasing the $\Val$ field of the installer's handle to $\X.\seq$ (Step 5).
Line~13, together with Lines 16 to 19, implement the second mission of forwarding the operation to $\Y$ (Step 6).
The if-condition and the CAS' arguments at Line~18 ensure that $\Y$ is changed only if $\Y.\seq$ lags behind $\X.\seq$ and, if it lags behind, it catches up and $\Y.\val$ takes on the installer's value.
The Recover method simply forwards at Line~20, as explained in Step 7.
The detect method returns at Line~22 the value in the handle's $\Val$ field, as explained in Step 8, along with $\true$ (since only successful ECSC operations are detected).
\\\\



\begin{footnotesize}
\begin{algorithm*}
\caption{: The $\DurEC$ class for Durable, External Context nW-LLSC objects.}
\label{alg:DurEC}
\begin{algorithmic}[1]

\medskip 

\Statex \cmdclass $\DurEC$:

\medskip 

\Statex \tab \cmdinstancevar \; (\Thandle, \Tint) \: $\X$ \Comment{$X$ is a pair $(\X.\hndl, \X.\seq)$ stored in NVM}
\Statex \tab \cmdinstancevar \; (\Tint, \Tint) \: $\Y$ \Comment{$Y$ is a pair $(\Y.\seq, \Y.\val)$ stored in NVM}

\medskip 

\Statex \tab \cmdstruct $\textit{handle} \:\{$
\Statex \tab[2] $\Tint \: \Last$
\Statex \tab[2] $\Tint \: \Val$
\Statex \tab $\}$

\medskip 

\Statex \tab \cmdstatic \cmdprocedure $\CreateHandle{}$
\State \tab[2] $\cmdreturn \: \cmdnew \: handle \{\Last = 0\}$ \Comment{fields $\Last$ and $\Val$ are stored in NVM; $\Val$ is arbitrarily initialized}

\medskip 

\Statex \tab \cmdconstructor $\DurEC(\Tint\: \initval)$
\State \tab[2] $\X \gets (null, 0)$
\State \tab[2] $\Y \gets (0, \initval)$

\medskip 

\Statex \tab \cmdprocedure $\CLL{\Thandle\: \h}$
\State \tab[2] \cmdreturn $\Y$

\medskip 

\Statex \tab \cmdprocedure $\CVL{\Thandle\: \h, \Tint \: \s}$
\State \tab[2] \cmdreturn $\Y.\seq = \s$

\medskip 

\Statex \tab \cmdprocedure $\CSC{\Thandle\: \h, \Tint \: \s, \Tint \: \v}$
\State \tab[2] \cmdif $\Y.\seq \ne \s$ \cmdthen \cmdreturn $\false$
\State \tab[2] $\h.\Val \gets \v$
\State \tab[2] $\hh \gets \X.\hndl$ 
\State \tab[2] $\sh \gets \max(\h.\Last, \s) + 1$
\State \tab[2] $\ret \gets \CAS{\X, (\hh, \s), (\h, \sh)}$
\State \tab[2] $\forward{\h}$
\State \tab[2] \cmdreturn $\ret$ 

\medskip 

\Statex \tab \cmdprocedure $\forward{\Thandle\: \h}$
\State \tab[2] $\x \gets \X$ 
\State \tab[2] $\sh \gets \x.\hndl.\Last$ 
\State \tab[2] \cmdif $\sh < \x.\seq$ \cmdthen $\CAS{\x.\hndl.\Last, \sh, \x.\seq}$

\State \tab[2] $\vh \gets \x.\hndl.\Val$ 
\State \tab[2] $\y \gets \Y$
\State \tab[2] \cmdif $\y.\seq < \x.\seq$ \cmdthen $\CAS{\Y, \y, (\x.\seq, \vh)}$
\State \tab[2] \cmdreturn

\medskip 

\Statex \tab \cmdprocedure $\Recover{\Thandle\: \h}$ 
\State \tab[2] $\forward{\h}$
\State \tab[2] \cmdreturn

\medskip 

\Statex \tab \cmdstatic \cmdprocedure $\Detect{\Thandle\: \h}$
\State \tab[2] \cmdreturn $\h.\Last$

\end{algorithmic}
\end{algorithm*}
\end{footnotesize}

\subsection{DurEC Proof Outline}

\noindent
The full proof of the $\DurEC$ algorithm is in Appendix~\ref{DurECProof}.
Here we reproduce the key definitions and lemmas.
\\\\
Let $\O$ be a DurEC object implemented by the algorithm, and $\X$ and $\Y$ be atomic CAS objects that $\O$ is implemented from.
The following two types of events are of interest.

\begin{itemize}
\item
An {\em install} is a successful CAS operation on $\X$, executed by a $\CSC{\h, \s, \v}$ operation $\alpha$ at Line~10.
We say $\alpha$ installs and $\alpha$ is an installer.
\item
A {\em move} is a successful CAS operation on $\Y$, executed by a $\forward{\h}$ operation $\alpha$ at Line~18.
We say $\alpha$ installs and $\alpha$ is a mover.
\end{itemize}






\begin{lemma}\label{all-durec-lemmas}
{\em
\mbox{ }

\begin{enumerate}
\item
Installs and moves alternate, starting with an install.
\item
If the latest event is a move or if no installs have occurred, then $\X.\seq = \Y.\seq$.
Otherwise (i.e., if the latest event is an install), $\X.\seq > \Y.\seq$.
\end{enumerate}
}
\end{lemma}

\begin{lemma}\label{tran1-durec-lemmas}
\emph{
If $\X.\seq > \Y.\seq$ at time $t$ and a $\forward{\h}$ operation $\alpha$ is started after $t$ and $\alpha$ completes without crashing, then a move occurs after $t$ and at or before $\alpha$'s completion time.
}
\end{lemma}

\begin{lemma}\label{wok-durec-lemmas}
\emph{
If a $\CSC{\h, \s, \v}$ operation $\alpha$ installs at time $t$, the first move after $t$ 
occurs by the time $\alpha$ completes.
}
\end{lemma}


\begin{lemma}\label{m1-durec-lemmas}
{\em
If a $\CSC{\h', \s, \v}$ operation $\alpha'$ installs at time $t'$ and a $\forward{\h}$ operation $\alpha$ moves at $t$ and is the first to move after $t'$, then:

\begin{enumerate}
\item
In the interval $(t', t)$, $\X.\hndl = \h$, $\h.\Val = \v$, and $\Y.\seq = \s$.
\item
$\alpha$ sets $\Y.\val$ to $\v$.
\end{enumerate}
}
\end{lemma}

We define a {\em hitchhiker} as a $\CSC$ operation that does not install and returns at Line~12.



\begin{lemma}\label{hh-durec-lemmas}
{\em
If $\alpha$ is a hitchhiker $\CSC$ operation, a move occurs during $\alpha$.
}
\end{lemma}

The next definition states how operations are linearized.
A crashed operation is not linearized, unless it is a $\CSC$ operation that crashes after installing.  
Hitchhikers return $\false$ at Line~12, so they are not crashed operations and are linearized.

\begin{definition}[Linearization]\label{wlin1-durec-lemmas}
{\em

\mbox{ }
\begin{enumerate}
\item
If a $\CSC{\h, \s, \v}$ operation $\alpha$ installs, it is linearized at the first move after $\alpha$'s install.

(Lemma~\ref{wok-durec-lemmas} guarantees that $\alpha$ is linearized before it completes.)

\item
If a $\CSC{\h, \s, \v}$ operation $\alpha$ is a hitchhiker, it is linearized at the earliest time $t$ such that 
a move occurs at $t$.
Furthermore, if $\beta$ is the installing $\CSC$ operation linearized at the same time $t$,
$\alpha$ is linearized {\em after} $\beta$.

{\em Remarks}: Lemma~\ref{hh-durec-lemmas} guarantees that $\alpha$ is linearized before it completes.
Linearizing a hitchhiker after the installer ensures that the success of the installer's ECSC causes the hitchhikers's ECSC to fail without changing the object's state, thereby eliminating the burden of detecting the hitchhikers' ECSC operation.

\item
If a $\CSC{\h, \s, \v}$ operation $\alpha$ returns at Line~6, it is linearized at Line 6.

\item
A $\CLL{\h}$ operation $\alpha$ is linearized at Line 4.

\item
A $\CVL{\h, \s}$ operation $\alpha$ is linearized at Line 5.
\end{enumerate}
}

\end{definition}

The value of a DurEC object implemented by the algorithm changes atomically at the linearization points of successful $\CSC$ operations.
The next lemma states that the algorithm maintains the DurEC object's state in $\Y$, and satisfies durable linearizability.

\begin{lemma}[Durable-linearizability of \DurEC \ objects]\label{juice-durec-lemmas}
{\em 
Let $\O$ be a DurEC object implemented by the algorithm.

\begin{enumerate}
\item
$(\O.\seq, \O.\val) = (\Y.\seq, \Y.\val)$ at all times.
\item
Let $\alpha$ be any $\O.\CSC{\h, \s, \v}$, $\O.\CLL{\h}$, or $\O.\CVL{\h}$ operation, and $t$ be the time at which $\alpha$ is linearized.
Suppose that $\O$'s state is $\sigma$ at $t$ just before $\alpha$'s linearization (in case multiple operations are linearized at $t$), and $\delta(\sigma, \alpha) = (\sigma', r)$, where $\delta$ is the sequential specification of a EC object. Then:

\begin{enumerate}
\item
$\O$'s state changes to $\sigma'$ at time $t$.
\item
If $\alpha$ completes without crashing, it returns $r$.

(Recall that if $\alpha$ crashes and, upon restart, executes $\Recover$, the recover method does not return any response.)
\end{enumerate}

\end{enumerate}
}
\end{lemma}

Next we state a key lemma for proving the detectability of  DurEC objects.

\begin{lemma}\label{det1-durec-lemmas}
{\em
\mbox{ }

\begin{enumerate}
\item
If a $\CSC{\h,\s,\v}$ operation $\alpha$ installs, then the value of $\h.\detval$ increases between  $\alpha$'s invocation and completion.
\item
For any handle $\h$, if $\h.\detval$ is changed at any time $t$ by the execution of Line~15 by some $\forward{\h'}$ method (for some $\h'$), then $\X.\hndl = \h$ and $\h.\pc \in \{13,14,15\}$.
\item
If a $\CSC{\h,\s,\v}$ operation $\alpha$ does not install, then the value of $\h.\detval$ is the same at $\alpha$'s invocation and completion.
\end{enumerate}
}
\end{lemma}

\begin{lemma}[Detectability of \DurEC \ objects]\label{det2-durec-lemmas}
{\em
Let $\alpha$ be any operation executed on a DurEC object $\O$ by a handle $\h$.
Suppose that $(d_1, r_1)$ and $(d_2, r_2)$ are the values that $\Detect{\h}$ would return, if executed immediately before $\alpha$ is invoked and immediately after $\alpha$ completes, respectively.
Then:
\begin{enumerate}
    \item
    If $\alpha$ is not an installing ECSC, it is safe to repeat and $d_2 = d_1$.
    \item
    If $\alpha$ is an installing ECSC, then $d_2 > d_1$ and $r_2 = \true$.
\end{enumerate}
}
\end{lemma}

\begin{restatable}{theorem}{durec}
Algorithm \DurEC \ satisfies the following properties:

\begin{enumerate}
\item
The objects implemented by the algorithm are durably linearizable (with respect to EC's sequential specification) and are detectable.
\item
All operations, including the Recover, Detect, Constructor, and CreateHandle methods, are wait-free and run in constant time.
\item
The algorithm supports dynamic joining: a new process can join in at any point in a run (by calling CreateHandle) and start creating DurEC objects or accessing existing DurEC objects.
\item
The space requirement is $O(m+n)$, where $m$ is the actual number of DurEC objects created in the run, and $n$ is the actual number of processes that have joined in in a run.
\end{enumerate}
\end{restatable}
 
A full proof of the $\DurEC$ algorithm is presented in \ref{DurECProof} (6 pages).


    
\section{$\DurECW$ and $\DuraLL$: durable Writable LLSC implementations}
\label{sec:WLLSC}

Using the {\em non-writable} $\DurEC$ building block of the previous section, we design the {\em writable} external context LLSC implementation $\DurECW$ in this section.
With $\DurECW$ in hand, we obtain our standard durable writable-LLSC implementation $\DuraLL$ easily, by simply rolling the context into the object.

\subsection{Intuitive description of Algorithm \DurECW}

A DurECW object $\O$ supports the write operation, besides  ECSC, for changing the object's state. 
Unlike a $\CSC{\h,\s,\v}$ operation, which returns without changing $\O$'s state when $\O.\context \neq \s$, a $\Write{\h, \v}$ must get $\v$ into $\O.\val$ unconditionally.
In  the DurECW algorithm, $\CSC$ operations help $\Write$ operations and prevent writes from being blocked by a continuous stream of successful $\CSC$ operations.

Each DurECW object $\O$ is implemented from two DurEC objects, $\W$ and $\Z$, each of which holds a pair, where the first component is a sequence number $\seq$, and the second component is a pair consisting of a value $\val$ and a bit $\bit$. Thus, $\W = (\W.\seq, (\W.\val, \W.\bit))$ and $\Z = (\Z.\seq, (\Z.\val, \Z.\bit))$.

\begin{footnotesize}
\begin{algorithm*}
\caption{
The $\DurECW$ class for Durable External Context W-LLSC objects. 
}
\label{alg:DurECW}

\begin{algorithmic}[1]

\Statex \cmdclass \DurECW:

\medskip 

\Statex \tab \textbf{instance variable} \; \textbf{DurEC} \: $\W$ 
\Comment{$\W$ holds a pair $(\W.seq, (\W.\val, \W.\bit))$}
\Statex \tab \textbf{instance variable} \; \textbf{DurEC} \: $\Z$ 
\Comment{$\Z$ holds a pair $(\Z.seq, (\Z.\val, \Z.\bit))$}

\medskip 

\Statex \tab $\textbf{struct} \: \textit{handle} \:\{$
\Statex \tab[2] $\textbf{DurEC.handle} \: \Critical$
\Statex \tab[2] $\textbf{DurEC.handle} \: \Casual$
\Statex \tab $\}$

\medskip 

\Statex \tab \cmdstatic \cmdprocedure $\CreateHandle{}$
\State \tab[1] \cmdreturn  $\textbf{handle} \; \{\Critical \gets \text{DurEC}.\CreateHandle{}, \, \Casual \gets \text{DurEC}.\CreateHandle{} \} $

\medskip 

\Statex \tab \cmdprocedure $\DurECW(\initval)$
\State \tab[2] $\W \gets \ECL{(0, 0)}$ 
\State \tab[2] $\Z \gets \ECL{(\initval, 0)}$

\medskip 

\Statex \tab \cmdprocedure $\CLL{\Thandle\: \h}$
\State \tab[2] $\z \gets \Z.\CLL{\h.\Casual}$
\State \tab[2] \cmdreturn $(\z.\seq, \z.\val)$

\medskip 

\Statex \tab \cmdprocedure $\CVL{\Thandle\: \h, \Tint \: \s}$
\State \tab[2] \cmdreturn $\Z.\CVL{\h.\Casual, \s}$

\medskip 

\Statex \tab \cmdprocedure $\CSC{\Thandle\: \h, \Tint \: \s, \Tint \: \v}$
\State \tab[2] $\z \gets \Z.\CLL{\h.\Casual}$
\State \tab[2] \cmdif $\s \ne \z.\seq$ \cmdreturn $\false$
\State \tab[2] $\transfer{\h}$
\State \tab[2] $\ret \gets \Z.\CSC{\h.\Critical, \s, (\v, \z.\bit)}$
\State \tab[2] \cmdreturn $\ret$

\medskip 

\Statex \tab \cmdprocedure $\Write{\Thandle\: \h, \Tint \: \v}$
\State \tab[2] $\w \gets \W.\CLL{\h.\Casual}$ 
\State \tab[2] $\z \gets \Z.\CLL{\h.\Casual}$
\State \tab[2] \cmdif $\z.\bit = \w.\bit$ \cmdthen $\W.\CSC{\h.\Critical, \w.\seq, (\v, 1 - \w.\bit)}$ 
\State \tab[2] $\transfer{\h}$
\State \tab[2] $\transfer{\h}$
\State \tab[2] \cmdreturn $\true$

\medskip 

\Statex \tab \cmdprocedure $\transfer{\Thandle \: \h}$
\State \tab[2] $\zh \gets \Z.\CLL{\h.\Casual}$
\State \tab[2] $\wh \gets \W.\CLL{\h.\Casual}$
\State \tab[2] \cmdif $\zh.\bit \neq \wh.\bit$ \cmdthen $\Z.\CSC{\h.\Casual, \zh.\seq, (\wh.\val, \wh.\bit)}$

\medskip 

\Statex \tab \cmdprocedure $\Recover{\Thandle \: \h}$
\State \tab[2] $\W.\Recover{\h.\Critical}$
\State \tab[2] $\Z.\Recover{\h.\Critical}$
\State \tab[2] $\W.\Recover{\h.\Casual}$
\State \tab[2] $\Z.\Recover{\h.\Casual}$
\State \tab[2] $\transfer{\h}$
\State \tab[2] $\transfer{\h}$

\medskip 

\Statex \tab \cmdstatic \cmdprocedure $\Detect{\textbf{handle*} \: \h}$
\State \tab[2] \cmdreturn $(\textsc{DurEC}.\Detect{\h.\Critical}, \true)$

\end{algorithmic}

\end{algorithm*}
\end{footnotesize}

The DurECW handle $\h$ consists of two DurEC handles, $\h.\Critical$ and $\h.\Casual$.
The use of two $\DurEC$ handles allows us to implement detectability.
In particular, if $\Detect{\h}$ is called on a $\DurECW$ object, only the detect value ($\Last$) of $\h.\Critical$ is returned.
So intuitively, when a DurECW operation $\alpha$ calls methods on $\W$ or $\Z$, it uses $\h.\Critical$ only if a successful call will make its own $\CSC$ or $\Write$ operation visible.
In all other cases $\alpha$ uses $\h.\Casual$.

The algorithm maintains the DurECW object $\O$'s state in $\Z$, i.e., $\O.\seq = \Z.\seq$ and $\O.\val = \Z.\val$ at all times. 
This representation makes the implementation of $\O.\CLL$ and $\O.\CVL$ operations obvious: $\O.\CLL{\h}$ simply returns $(\Z.\seq, \Z.\val)$ and $\CVL{\h, \s}$ returns whether $\Z.\seq = \s$.
The complexity lies in the implementation of $\O.\Write{\h,\v}$ and $\O.\CSC{\h, \s, \v}$ operations, which coordinate their actions using $\W.\bit$ and $\Z.\bit$. A write operation flips the $\W.\bit$ to announce to the ECSC operations that their help is needed to push the write into $\Z$; once the write is helped, the $\Z.\bit$ is flipped to announce that help is no longer needed. 
We maintain the invariant that $\W.\bit \neq \Z.bit$ if and only if a write needs help. 

A $\Write{\h, \v}$ operation $\alpha$ consists of the following steps.

\begin{enumerate}
\item[(W1)]
The operation $\alpha$ reads $\W$ and $\Z$ to determine if some write operation is already waiting for help.
If not, then $\alpha$ installs its write into $\W$ by setting $\W.\val$ to $\v$ and flipping $\W.\bit$.
If several write operations attempt to install concurrently, only one will succeed.
The successful one is the {\em installer} and the others are {\em hitchhikers}.
\item[(W2)]
Once a write operation is installed, all processes---installer, hitchhiker, and the ECSC operations---work in concert to forward the installer's operation to $\Z$.
Since $\Z$ is where the DurECW object's state is held, the installer's operation takes effect only when it is reflected in $\Z$'s state.
Towards this end, everyone attempts to transfer the installer's value from $\W$ to $\Z$. However, a stale ECSC operation, which was poised to execute its ECSC operation on $\Z$, might update $\Z$, causing the transfer to fail in moving the installer's value from $\W$ to $\Z$.
So, a transfer is attempted the second time.
The earlier success by the poised ECSC operation causes any future attempts by similarly poised operations to fail.
Consequently, the installer's write value gets moved to $\Z$ by the time the second transfer attempt completes.
The point where the move to $\Z$ occurs becomes the linearization point for the installer's write operation.
We linearize the writes by the hitchhikers immediately before the installer, which makes their write operations to be overwritten immediately by the installer's write, without anyone ever witnessing their writes.
Hence, there is no need to detect these writes: if a hitchhiker crashes during its write, the operation can be safely repeated.

\item[(W3)]
If the installer crashes after installing, upon restart, in the Recover method, it does the forwarding so that its install moves to $\Z$ and its write operation gets linearized.

\end{enumerate}

An $\CSC{\h, \s, \v}$ operation $\alpha$ consists of the following steps.
\begin{enumerate}
    \item[(S1)] 
    $\alpha$ performs an $\CLL$ to determine whether the context in $\O$ matches $\s$.  If not, it can fail early and return $\false$.
    \item[(S2)]
    If a $\Write$ is already in $\W$ and waiting for help to be transferred to $\Z$, $\alpha$ is obligated to help that write before attempting its SC (to prevent the write from being blocked by a chain of successful $\CSC$ operations). So it attempts a transfer from $\W$ to $\Z$.
    \item[(S3)] 
    Finally $\alpha$ executes an $\CSC$ on $\Z$ in an attempt to make its own operation $\O$ take effect.
\end{enumerate}

The algorithm is formally presented in \ref{alg:DurECW}. In the algorithm, Lines 12-14 implement step W1 and Lines 15, 16 implement step W2. Step S1 is implemented by Lines 7, 8, step S2 by 9 and S3 by 10 and 11. Note that the $\CSC$ on line 10 takes care to not change $\Z.\bit$. This ensures that the helping mechanism for writes implemented via $\W.\bit$ and $\Z.\bit$ is not disturbed.
The $\CSC$ operation at Line 14 uses the handle $\h.\Critical$ because its success implies that the operation is an installer and hence will be a visible write when it linearizes.  
Similarly the $\CSC$ on $\Z$ at Line 10 uses $\h.\Critical$ because its success makes the $\CSC$ on $\O$ visible.

If a $\Write$ or a $\CSC$ method crashes while executing an operation on $\W$ or $\Z$, upon restart, Lines 21 to 24 of $\Recover$ ensure that $\W.\Recover$ or $\Z.\Recover$ is executed before any other operation is executed on $\W$ or $\Z$.  Consequently, the durable objects $\W$ and $\Z$ behave like atomic EC objects. 
\\\\
The theorem below summarizes the result:

\begin{restatable}{theorem}{durlindurecw}
Algorithm \DurECW \ satisfies the following properties:

\begin{enumerate}
\item
The objects implemented by the algorithm are durably linearizable (with respect to ECW's sequential specification) and are detectable.
\item
All operations, including the Recover, Detect, Constructor, and CreateHandle methods, are wait-free and run in constant time.
\item
The algorithm supports dynamic joining: a new process can join in at any point in a run (by calling CreateHandle) and start creating objects or accessing existing objects.
\item
The space requirement is $O(m+n)$, where $m$ is the actual number of DurECW objects created in the run, and $n$ is the actual number of processes that have joined in in a run.
\end{enumerate}
\end{restatable}

{\em Proof}: A detailed proof of this theorem is presented in \ref{DurECWProof} (7 pages).

    \subsection{The $\DuraLL$ Algorithm}
\label{sec:DuraLL}

Given the durable EC W-LLSC object $\DurECW$, rolling the context into the implementation to produce a durable standard W-LLSC object is simple.
Each of our implemented $\DuraLL$ objects simply maintains a single $\DurECW$ object $\X$.
The handle of the $\DuraLL$ object simply maintains a single $\DurECW$ handle, to operate on $\X$, and a hashmap that maps objects to $\contexts$.

\begin{footnotesize}
\begin{algorithm*}
\caption{
The $\DuraLL$ class for Durable Writable-LLSC objects. 
}
\label{alg:DuraLL}

\begin{algorithmic}[1]

\Statex \cmdclass \DuraLL:

\medskip 

\Statex \tab \textbf{instance variable} \; \textbf{DurECW} \: $\X$ 
\Comment{$\X$ holds the central EC W-LLSC object.}

\medskip 

\Statex \tab $\cmdstruct \: \textit{handle} \:\{$
\Statex \tab[2] $\textbf{DurECW.handle} \: \ECWH$
\Statex \tab[2] $\textbf{HashMap} \: (\textbf{DuraLL} \rightarrow \textbf{int})  \: \contexts$
\Statex \tab $\}$

\medskip

\Statex \tab \cmdstatic \cmdprocedure $\CreateHandle{}$
\State \tab[2] \cmdreturn  $\textbf{handle}$ \{$\ECWH \gets \DurECW.\CreateHandle{},$
$\contexts \gets \textbf{HashMap} \: (\textbf{DuraLL} \rightarrow \textbf{int}) \} $

\medskip 

\Statex \tab \cmdprocedure $\DuraLL(\initval)$
\State \tab[2] $\X \gets \DurECW{(\initval, 0)}$ 

\medskip 

\Statex \tab \cmdprocedure $\LL{\Thandle \: \h}$
\State \tab[2] $\x \gets \X.\CLL{\h.\Critical}$
\State \tab[2] $\h.\contexts(\self) \gets \x.\seq$
\State \tab[2] \cmdreturn $\x.\val$

\medskip 

\Statex \tab \cmdprocedure $\VL{\Thandle \: \h}$
\State \tab[2] \cmdif $\self \not\in \h.\contexts.keys$ \cmdthen \cmdreturn $\false$
\State \tab[2] \cmdreturn $\X.\CVL{\h.\ECWH, \h.\contexts(\self)}$

\medskip 

\Statex \tab \cmdprocedure $\SC{\Thandle \: \h, \Tint \: \val}$
\State \tab[2] \cmdif $\self \not\in \h.\contexts.keys$ \cmdthen \cmdreturn $\false$
\State \tab[2] $\ret \gets \X.\CSC{\h.\ECWH, \h.\contexts(\self), \val}$
\State \tab[2] $\h.\contexts.\Remove{\self}$
\State \tab[2] \cmdreturn $\ret$

\medskip 

\Statex \tab \cmdprocedure $\Write{\Thandle \: \h,\Tint \: \val}$
\State \tab[2] $\X.\Write{\h.\ECWH, \val}$ 
\State \tab[2] $\h.\contexts.\Remove{\self}$
\State \tab[2] \cmdreturn $\true$

\medskip 

\Statex \tab \cmdprocedure $\Recover{\Thandle \: \h}$
\State \tab[2] $\X.\Recover{\h.\ECWH}$
\State \tab[2] \cmdif $\self \in \h.\contexts.keys$ \cmdthen \cmdif $\neg \X.\CVL{\h.\ECWH, \h.\contexts(\self)}$ \cmdthen $\h.\context.\Remove{\self}$

\medskip 

\Statex \tab \cmdstatic \cmdprocedure $\Detect{\Thandle \: \h}$
\State \tab[2] \cmdreturn $\DurECW.\Detect{\h.\ECWH}$

\end{algorithmic}

\end{algorithm*}
\end{footnotesize}

We present the code as Algorithm~\ref{alg:DuraLL}.
The $\LL$ operation on a $\DuraLL$ object by handle $\h$ simply performs a $\CLL$ on $\X$ and stores the returned context in $\h.\contexts$ under the key $\self$ (which is the reference of the current object).
Correspondingly, $\VL$ retrieves the context from $\h.\contexts$, and uses it to perform a $\CVL$ on $\X$.
The $\SC$ operation also retrieves the context and performs a $\CSC$ on the internal object, but then cleverly removes the key corresponding to the current object from $\h.\contexts$, since, regardless of whether the $\SC$ succeeds, the stored context is bound to be out-of-date.
The $\Write$ operation does not need a context, so it simply writes to $\X$, but also cleverly removes the current object's key from $\h.\contexts$ to save some space.
In order to be space-efficient, $\Recover$ also removes the current object from $\h.\contexts$ if the context stored for the object is out-of-date.
Since $\DuraLL$ is just a wrapper around $\DurECW$, its $\Detect$ operation simply returns the result of detecting $\DurECW$.

\begin{theorem}
Algorithm $\DuraLL$ \ satisfies the following properties:
\begin{enumerate}
\item
The objects implemented by the algorithm are durably linearizable (with respect to LL/SC's sequential specification) and are detectable.
\item
All operations, including the Recover, Detect, Constructor, and CreateHandle methods, are wait-free and run in constant time.
\item
The algorithm supports dynamic joining: a new process can join in at any point in a run (by calling CreateHandle) and start creating DuraLL objects or accessing existing DuraLL objects.
\item
The space requirement is $O(m+n+C)$, where $m$ is the actual number of DuraLL objects created in the run, $n$ is the actual number of processes that have joined in in a run, and $C$ is the number of ``contexts" stored across all objects.
\end{enumerate}
\end{theorem}

\section{$\DuraCAS$: a durable implementation of Writable CAS}
\label{sec:WCAS}

Using the $\DurEC$ building block, we design a Writable-CAS object, $\DuraCAS$.
The $\DuraCAS$ algorithm resembles $\DurECW$, but requires some new ideas due to the subtle differences between LLSC and CAS. 

\subsection{Informal description of Algorithm \DuraCAS}

We present in Figure~\ref{alg:DuraCAS} Algorithm \DuraCAS, which implements a durable writable CAS object $\O$ from two DurEC objects, $\W$ and $\Z$. 
The algorithm bears a lot of similarity to Algorithm $\DurECW$ of the previous section.
In fact, $\DuraCAS$ has only three extra lines.
For readability, we starred their line numbers (Lines {\bf 6*}, {\bf 10*}, and {\bf 13*}) and kept the line numbers the same for the common lines.

The ideas underlying this algorithm are similar to $\DurECW$, so we explain here only the three differences:
(1) Lines 7 to 10 are executed only once in Algorithm \DurECW, but are repeated twice in the current algorithm;
(2) Line~8 differs in the two algorithms; and
(3) Line~13* is introduced in the current algorithm.

The change in Line~8 accounts for the fact that the success of a $\CAS$ operation depends on the value in $\O$ rather than the context.  If the value in $\O$ (and therefore $\Z$) is different from $\old$ at Line 7, the CAS returns $\false$ (and linearizes at Line 7).  If $\O.\val = \old$ and the CAS does not plan to change the value (i.e., $\old = \new$) it returns $\true$ without changing $\Z$.

To understand why Lines 7 to 10 are repeated in the current algorithm, consider the following scenario.
A handle $\h$ executes $\O.\cas(\h, \old, \new)$, where $\old \neq \new$.
When $\h$ executes Line~7, $\Z$'s value is $\old$, so $\z.\val$ gets set to $\old$ at Line~7.
Handle $\h$ progresses to Line~10, but before it executes Line~10, some handle $\h'$ invokes $\O.\Write{\h', \old}$ and executes it to completion, causing $\Z.\seq$ to take on a value greater than $\z.\seq$.
Handle $\h$ now executes the ECSC at Line~10
and fails since $\Z.\seq \neq \z.\seq$.
If $\h$ acts as it did in Algorithm \DurECW, $\h$ would complete its $\O.\cas(\h, \old, \new)$ operation, returning $\false$.
However, $\false$ is an incorrect response by the specification of CAS because $\O.\val = \old$ for the full duration of the operation $\O.\cas(\h, \old, \new)$.
To overcome this race condition, $\h$ repeats Lines 7 to 10.

\begin{footnotesize}
\begin{algorithm*}
\caption{
The $\DuraCAS$ class for Durable, Writable-CAS objects. 
}
\label{alg:DuraCAS}

\begin{algorithmic}[1]

\Statex \cmdclass \DuraCAS:

\medskip 

\Statex \tab \cmdinstancevar \; \textbf{DurEC} \: $\W$ 
\Comment{$\W$ holds a pair $(\W.\seq, (\W.\val, \W.\bit))$}
\Statex \tab \cmdinstancevar \; \textbf{DurEC} \: $\Z$ 
\Comment{$\Z$ holds a pair $(\Z.\seq, (\Z.\val, \Z.\bit))$} 

\medskip 

\Statex \tab $\cmdstruct \: \textit{handle} \:\{$
\Statex \tab[2] $\textbf{DurEC}.\Thandle \: \Critical$
\Statex \tab[2] $\textbf{DurEC}.\Thandle \: \Casual$
\Statex \tab $\}$

\medskip 

\Statex \tab \cmdstatic \cmdprocedure $\CreateHandle{}$
\State \tab[2] \cmdreturn \cmdnew 
$handle \{
\Critical \gets \DurEC.\CreateHandle{},  
\Casual \gets \DurEC.\CreateHandle{}
\}$

\medskip 

\Statex \tab \cmdprocedure $\DuraCAS(\Tint \: \initval)$
\State \tab[2] $\W \gets \ECL{(0, 0)}$ 
\State \tab[2] $\Z \gets \ECL{(\initval, 0)}$

\medskip 

\Statex \tab \cmdprocedure $\Read{\Thandle \: \h}$
\State \tab[2] $\z \gets \Z.\CLL{\h.\Casual}$
\State \tab[2] \cmdreturn $\z.\val$

\medskip 

\State 

\medskip 

\Statex \tab \cmdprocedure $\Rcas{\Thandle \: \h, \Tint \: \old, \Tint \: \new}$

\Statex \mbox{\hspace{-\algorithmicindent}} {\footnotesize {\bf 6*}:}\hspace{\algorithmicindent} \tab \cmdfor $i \gets 1 \mbox{ to } 2$
\State \tab[3] $\z \gets \Z.\CLL{\h.\Casual}$ 
\State \tab[3] \cmdif $\z.\val \ne \old$ \cmdthen \cmdreturn $\false$ \cmdelse \cmdif $\old = \new$ \cmdthen \cmdreturn $\true$
\State \tab[3] $\transfer{\h}$
\State \tab[3] \cmdif $\Z.\CSC{\h.\Critical, \z.\seq, (\new, \z.bit)}$ \cmdthen
\Statex \mbox{\hspace{-0.25in}} {\small {\bf 10*}:}\mbox{\hspace{0.165in}} \tab[3] \cmdreturn $\true$ 
\State \tab[2] \cmdreturn $\false$

\medskip 

\Statex \tab \cmdprocedure $\Write{\Thandle \: \h,\Tint \: \v}$
\State \tab[2] $\w \gets \W.\CLL{\h.\Casual}$ 
\State \tab[2] $\z \gets \Z.\CLL{\h.\Casual}$
\Statex \mbox{\hspace{-0.25in}} {\small {\bf 13*}:}\mbox{\hspace{0.165in}} \tab[1] \cmdif $\z.\val = \v$ \cmdthen \cmdreturn $ack$
\State \tab[2] \cmdif $\z.\bit = \w.\bit$ \cmdthen $\W.\CSC{\h.\Critical, \w.\seq, (\v, 1 - \w.\bit)}$ 
\State \tab[2] $\transfer{\h}$ 
\State \tab[2] $\transfer{\h}$ 
\State \tab[2] \cmdreturn $ack$

\medskip 

\Statex \tab \cmdprocedure $\transfer{\Thandle \: \h}$
\State \tab[2] $\zh \gets \Z.\CLL{\h.\Casual}$
\State \tab[2] $\wh \gets \W.\CLL{\h.\Casual}$
\State \tab[2] \cmdif $\zh.\bit \neq \wh.\bit$ \cmdthen $\Z.\CSC{\h.\Casual, \zh.\seq, (\wh.\val, \wh.\bit)}$

\medskip 

\Statex \tab \cmdprocedure $\Recover{\Thandle \: \h}$
\State \tab[2] $\W.\Recover{\h.\Critical}$
\State \tab[2] $\Z.\Recover{\h.\Critical}$
\State \tab[2] $\W.\Recover{\h.\Casual}$
\State \tab[2] $\Z.\Recover{\h.\Casual}$
\State \tab[2] $\transfer{\h}$
\State \tab[2] $\transfer{\h}$

\medskip 

\Statex \tab \cmdstatic \cmdprocedure $\Detect{\Thandle \: \h}$
\State \tab[2] \cmdreturn $(\textsc{DurEC}.\Detect{\h.\Critical}, \true)$

\end{algorithmic}

\end{algorithm*}
\end{footnotesize}

If the same race condition repeats each time $\h$ repeats Lines 7 to 10, the method $\O.\cas$ would not be wait-free.
Line~13* is introduced precisely to prevent this adverse possibility.
When a handle $\h'$ executes Lines 12 to 14 of $\O.\Write{\h', \v}$  in the previous $\DurECW$ algorithm, $\h'$ would always try to install its value $v$ in $\W$ (at Line~14) and later move it to $\Z$, thereby increasing $\Z.\seq$ and causing concurrent $\O.\CSC$ operations to fail.
This was precisely what we wanted because the specification of an SC operation requires that if {\em any} $\O.\Write$ takes effect, regardless of what value it writes in $\O$, it must change $\O.\context$ and thus cause concurrent $\O.\CSC$ operations to fail.
The situation however, is different when implementing $\O.\cas$, where a $\O.\Write$ that does not change the value in $\O$ should not cause a concurrent $\O.\cas$ to fail.
Hence, if a $\O.\Write{\h',v}$ operation is writing the same value as $\O$'s current value, then it should simply return (since $\O.\val$ already has $v$) and, importantly, not change $\Z.\seq$ (because changing $\Z.\seq$ would cause any concurrent $\cas$ operation to fail).
Line~13* implements precisely this insight.
\\\\
The theorem below summarizes the result:

\begin{restatable}{theorem}{durlin}
Algorithm \DuraCAS \ satisfies the following properties:

\begin{enumerate}
\item
The objects implemented by the algorithm are durably linearizable (with respect to the sequential specification of Writable CAS) and are detectable.
\item
All operations, including the Recover, Detect, Constructor, and CreateHandle methods, are wait-free and run in constant time.
\item
The algorithm supports dynamic joining: a new process can join in at any point in a run (by calling CreateHandle) and start creating objects or accessing existing objects.
\item
The space requirement is $O(m+n)$, where $m$ is the actual number of DuraCAS objects created in the run, and $n$ is the actual number of processes that have joined in in a run.
\end{enumerate}
\end{restatable}

{\em Proof}: A detailed proof of this theorem is presented in Appendix \ref{DuraCasProof} (8 pages).
\hfill $\blacksquare$


\section{Discussion and Remarks}
\label{sec:Conclusion}

In this paper, we have designed constant time implementations for durable CAS and LLSC objects.
To our knowledge, $\DuraCAS$ is the first CAS implementation to allow for dynamic joining.
$\DuraCAS$ also has state-of-the-art space complexity---allowing adaptivity and requiring only constant space per object and per process that actually accesses the protocol---and is writable.
To our knowledge, ours are the first implementations of durable LLSC objects. 
LLSC objects are universal and ABA-free, thus we believe that the dynamically joinable LLSC implementations in this paper will be useful in the construction of several more complex durable objects.
The external context variant of LLSC is particularly space efficient, making it a powerful building block for concurrent algorithms;
we witnessed this property even in the constructions of this paper, where the EC nW-LLSC object $\DurEC$ served as the primary building block for all our other implementations, including our EC W-LLSC implementation $\DurECW$ and its direct descendent $\DuraLL$ (for W-LLSC). 
All the implementations in this paper were enabled by handles---a novel, pointer-based mechanism we introduced in this paper to enable threads created on-the-fly to access our implementations.
We believe that along with the specific implementations of this paper, the use of handles as an algorithmic tool can play an important role in the design of future durable algorithms. 

We end with two open problems.
Handles enable dynamic joining, but once a handle $h$ is used, any other process can have a stale pointer to $h$ that may be dereferenced at any point in the future.
A mechanism for enabling space adaptivity for both dynamic joining and {\em dynamic leaving}, which would enable a process to reclaim its entire memory footprint once it is done using a durable implementation is our first open problem.
Our second open problem is to prove (or disprove) an $\Omega(m + n)$ space lower bound for supporting $m$ objects for $n$ processes for any durable CAS or durable LLSC type.


\clearpage
\bibliographystyle{acm}

\appendix
\section{Appendix}
\subsection{Proof of correctness of {\sc DurEC}}
\label{DurECProof}

Let $\O$ be a DurEC object implemented by the algorithm, and $\X$ and $\Y$ be atomic CAS objects that $\O$ is implemented from.
The following two types of events are of interest.

\begin{itemize}
\item
An {\em install} is a successful CAS operation on $\X$, executed by a $\CSC{\h, \s, \v}$ operation $\alpha$ at Line~10.
We say $\alpha$ installs and $\alpha$ is an installer.
\item
A {\em move} is a successful CAS operation on $\Y$, executed by a $\forward{\h}$ operation $\alpha$ at Line~18.
We say $\alpha$ installs and $\alpha$ is a mover.
\end{itemize}


To refer to the times when various actions are performed, we adopt the following notation.
If $\alpha$ is an execution of a $\CSC$, or $\forward$ and $\ell$ is a line number in the algorithm, $\alpha[\ell]$ denotes the time at which $\alpha$ executes Line $\ell$.
We denote an open interval from time $t_1$ to $t_2$ by $(t_1,t_2)$, the closed interval by $[t_1,t_2]$, and the semi-closed interval that includes $t_1$ and not $t_2$ by $[t_1,t_2)$.

\begin{lemma}\label{prelim1}
{\em
\mbox{ }

\begin{enumerate}
\item
$\X$ changes only at installs, and $\X.\seq$ increases at every install.
When a $\CSC{\h,\s,\v}$ operation installs, $\X.\seq$ increases from $\s$.
\item
$\Y$ changes only at moves, and $\Y.\seq$ increases at every move.
\end{enumerate}
}
\end{lemma}

\noindent
{\em Proof}: Line~10 is the only line where an attempt is made to change $\X$.
If the CAS at Line~10 succeeds, it follows from the arguments of the CAS and the setting of $\sh$ at Line~9 that $\sh > \s$ and $\X.\seq$ increases from $\s$ to $\sh$.
Hence, we have Part (1).

Line~18 is the only line where an attempt is made to change $\Y$.
If the CAS at Line~18 succeeds, it follows from the arguments of the CAS and the if-condition at Line~18 that $\Y.\seq$ increases.
Hence, we have Part (2).
\hfill $\blacksquare$

The next lemma states that install and move events alternate, starting with an install. 
An install causes $\X.\seq$ to exceed $\Y.\seq$ and a move brings up $\Y.\seq$ to equal $\X.\seq$.

\begin{lemma}\label{all}
{\em
\mbox{ }

\begin{enumerate}
\item
Installs and moves alternate, starting with an install.
\item
If the latest event is a move or if no installs have occurred, then $\X.\seq = \Y.\seq$.
Otherwise (i.e., if the latest event is an install), $\X.\seq > \Y.\seq$.
\end{enumerate}
}
\end{lemma}

\noindent
{\em Proof}:
We prove the lemma by induction.
For the base case, we show that the first event must be an install and $\X.\seq = \Y.\seq$ until the first install occurs.
Assume for a contradiction that a move occurs before any install occurs.
Let $\alpha$ be the $\forward{\h}$ operation that performs the first move, by executing a successful CAS at its Line~18.
Since no installs or moves have occurred before this move, the value $\x$ that $\alpha$ reads from $\X$ at Line~13 and the value $\y$ that $\alpha$ reads from $\Y$ at Line~17 are the initial values of $\X$ and $\Y$, respectively.
Since $\X.\seq$ and $\Y.\seq$ are both initially 0, it follows that $\x.\seq = \y.\seq$ when $\alpha$ executes Line~18.
Therefore, the if-condition at Line~18 prevents $\alpha$ from performing the CAS at Line~18, contradicting that $\alpha$ performs a move.
We conclude that the first event must be an install and $\X.\seq = \Y.\seq = 0$ until it occurs.
Hence, we have the base case of the induction.

For the induction step, we assume that $k-1 \ge 1$ events have occurred before time $t$, the lemma has held until $t$, and the $k$th event occurs at $t$.
Let $t'$ be the latest time before $t$ when an install or a move occurred.
Since no install or move events occur during the open interval $(t', t)$, the values of $X$ and $Y$ do not change during this interval.
Let $a$ and $b$ be the values of $\X.\seq$ and $\Y.\seq$, respectively, during $(t', t)$.
We prove the following four claims to complete the induction step.

\begin{itemize}
\item
\underline{Claim 1}: 
If an install occurred at $t'$, then an install cannot occur at $t$.

Assume for a contraction that the event at $t$ is an install, executed by a $\CSC{\h, \s, \v}$ operation $\alpha$.
Then, $t = \alpha[10]$.
By the induction hypothesis, $a > b$ and, at $\alpha[6]$, $\Y.\seq \le b$.
Furthermore, at $\alpha[6]$, $\Y.\seq$ must be $\s$; otherwise $\alpha$ would have returned at Line~8.
Putting these observations together, we have $\s \le b < a$; in particular, $\s < a$.
This, together with the fact that $\X.\seq = a$ when $\alpha$ executes the CAS at Line~10, implies that the CAS fails, contradicting that $\alpha$ installs.

\item
\underline{Claim 2}: 
If a move occurred at $t'$, then a move cannot occur at $t$.

Assume for a contraction that the event at $t$ is a move, executed by a $\forward{\h}$ operation $\alpha$.
Then, $t = \alpha[18]$.
By the induction hypothesis, $a = b$ during the interval $(t', t)$ and $\Y.\seq < b$ before $t'$.
Since $\alpha$'s CAS at Line~18 is successful,
from the arguments of the CAS operation there, it is clear that $\y = \Y$ just before $\alpha$ performs the CAS at $\alpha[18]$.
Since $\Y.seq = b$ just before $\alpha[18]$, it follows that $\y.\seq = b$.
Since $\X.\seq = a$ at $\alpha[18]$, it follows that $\X.\seq \le a$ at $\alpha[13]$, when $\alpha$ reads $\X$ into $\x$.
Since $a=b$, it follows that $\x.\seq \le a = b$.
Putting the above observations together, 
when $\alpha$ executes Line~18, we have $\x.\seq \le b = \y.\seq$.
Consequently, the if-condition at Line~18 prevents $\alpha$ from performing the CAS at Line~18, contradicting that $\alpha$ moves.

\item
\underline{Claim 3}: If a move occurred at $t'$ and an install occurs at $t$, then the install at $t$ increases $\X.\seq$.

For a proof of this claim, observe that $\sh$ is set to be greater than $\s$ at the earlier line (Line~9), and the install by the successful CAS at Line~10 increases $\X.\seq$ from $\s$ to $\sh$.

\item
\underline{Claim 4}: If an install occurred at $t'$ and a move occurs at $t$, then the move at $t$ increases $\Y.\seq$ and makes it equal to $\X.\seq$.
(It is obvious from Line~18 that every move increases $\Y.\seq$, but it is not obvious that the increase makes $\Y.\seq$ equal to $\X.\seq$, as claimed.)

For a proof of this claim, let $\alpha$ denote the $\forward{\h}$ operation that moves at $t$, which implies that $t = \alpha[18]$.
Let $t''$ be the time of the latest move before $t'$; if there is no move before $t'$, let $t'' = 0$.
By the induction hypothesis, $a > b$; $\Y.\seq = b$ in the interval $(t'', t)$; and $\X.\seq \le b$ in the interval $(0, t')$.

We assert that $\alpha[17] > t''$.
If this assertion were false, because of the move at $t''$, $\Y$'s state changes between $\alpha[17]$ and $\alpha[18]$, which implies that $\y \neq \Y$ at $\alpha$'s Line~18.
Hence, $\alpha$'s CAS fails at Line~18, contradicting that $\alpha$ moves.

We also assert that $\alpha[13] > t'$.
If this assertion were false, when $\alpha$ reads $\X$ into $\x$ at Line~13, the earlier inequalities imply that $\x.\seq \le b$.
From the previous assertion that $\alpha[17] > t''$ and the earlier inequalities, when $\alpha$ reads $\Y$ into $\y$ at Line~17, we have $\y.\seq = b$.
Putting these observations together, we have $\x.\seq \le b = \y.\seq$.
Therefore, when $\alpha$ executes Line~18, the if-condition there evaluates to $\false$ and $\alpha$ does not execute the CAS at that line, contradicting that $\alpha$ moves.

It follows from the previous assertion that $t' < \alpha[13] < \alpha[17] < \alpha[18]$.
So, when $\alpha$ reads $\X$ into $\x$ at Line~13 and $\Y$ into $\y$ at Line~17, the following holds when $\alpha$ performs the CAS at Line~18: $\X.\seq = \x.\seq = a > b = \y.\seq = \Y.seq$.
The success of that CAS increases $\Y.\seq$ from $b$ to $a$, thereby making $\X.\seq = \Y.\seq$.
\end{itemize}

Hence, we have the lemma.
\hfill $\blacksquare$

\begin{lemma}\label{tran1}
\emph{
If $\X.\seq > \Y.\seq$ at time $t$ and a $\forward{\h}$ operation $\alpha$ is started after $t$ and $\alpha$ completes without crashing, then a move occurs after $t$ and at or before $\alpha$'s completion time of $\alpha[19]$.
}
\end{lemma}

\noindent{\em Proof}: 
Assume to the contrary that no move occurs between $t$ and $\alpha$'s completion.
Then, Lemmas \ref{prelim1} and \ref{all} imply that the states of $\X$ and $\Y$ are unchanged during this interval $(t, \alpha[19])$.
Let $a$ and $b$ be the values of $\X$ and $\Y$ during this interval;
from the premise of the lemma, we have $a.\seq > b.\seq$.
When $\alpha$ reads $\X$ into $\x$ at Line~13 and $\Y$ into $\y$ at Line~17,
we have $\x = a$ and $\y = b$.
Therefore, the if-condition at Line~18 evaluates to $\true$, and the CAS at Line~18 succeeds.
Thus, a move occurs by $\alpha[19]$.
\hfill $\blacksquare$

\begin{lemma}\label{wok-durec}
\emph{
If a $\CSC{\h, \s, \v}$ operation $\alpha$ installs at time $t$, the first move after $t$ 
occurs by the time $\alpha$ completes.
}
\end{lemma}

\noindent{\em Proof}: After $\alpha$ installs at time $t$ (by executing a successful CAS at Line~10), $\alpha$ executes the $\forward$ method at Line~11.
If the process $p$ executing $\alpha$ crashes before completing the $\forward$ method, upon restart, $p$ executes the $\Recover{\h}$ method, which executes $\forward{\h}$ at Line~20.
Thus, after $\alpha$ installs at time $t$, regardless of crashes, $\alpha$ executes $\forward$ to completion  before its own completion.
It follows from Lemma~\ref{tran1} that a move occurs before $\alpha$ is completed.
Hence, we have the lemma.
\hfill $\blacksquare$ 

The next lemma states that if $i$ is an install event and $m$ is the earliest move event after $i$, the mover of $m$ must start after $i$.

\begin{lemma}\label{m1}
{\em
If a $\CSC{\h', \s, \v}$ operation $\alpha'$ installs at time $t'$ and a $\forward{\h}$ operation $\alpha$ moves at $t$ and is the first to move after $t'$, then:

\begin{enumerate}
\item
In the interval $(t', t)$, $\X.\hndl = \h$, $\h.\Val = \v$, and $\Y.\seq = \s$.
\item
$\alpha[13] > t'$
\item
$\alpha$ sets $\Y.\val$ to $\v$.
\end{enumerate}
}
\end{lemma}

\noindent
\underline{\em Proof of Part (1)}: 
When $\alpha'$ installs at $t'$, it sets $\X.\hndl$ to $\h$, and $\X$ is not changed until the next install, which is after the move at $t$ (by Lemmas~\ref{prelim1} and \ref{all}).
Hence, we have Part (1a).

$\h.\Val$ is set to $\v$ at Line~7 by the handle $\h$, and $\h.\Val$ is not changed by any other line in the algorithm.
So, $\h.\Val$ holds $\v$ from time $\alpha[3]$ to the time that handle $\h$ executes Line~3 in a later $\CSC$ operation (possibly on a different \DurEC \ object).
This, together with the fact the move at $t$ occurs before $\alpha$ completes (Lemma~\ref{wok-durec}), implies Part (1b).

Since $\alpha$ installs at $t'$, its CAS on $\X$ at $t'$ succeeds.
Therefore, from the arguments of that CAS at Line~10, it follows that $\X.\seq = \s$ just before the install.
Then, it follows from Lemma~\ref{prelim1} that $\Y.\seq = \s$ from the move preceding the install at $t'$ to the move at $t$.
Hence, we have Part (1c).

\vspace{0.1in}
\noindent
\underline{\em Proof of Part (2)}: 
Since $\alpha$ moves at $t$,  $t = \alpha[18]$.
Let $t''$ be the time of the latest move before $t'$; if there is no move before $t'$, let $t'' = 0$.
It follows from Lemma~\ref{all} that there are integers $a$ and $b$ such that $a>b$,
$\Y.\seq = b$ in the interval $(t'', t)$, $\X.\seq = a$ in the interval $(t', t)$, and $\X.\seq \le b$ in the interval $(0, t')$.

We assert that $\alpha[17] > t''$.
If this assertion were false, because of the move at $t''$, $\Y$'s state changes between $\alpha[17]$ and $\alpha[18]$, which implies that $\y \neq \Y$ at $\alpha$'s Line~18.
Hence, $\alpha$'s CAS fails at Line~18, contradicting that $\alpha$ moves.

Assume, contrary to the lemma, that  $\alpha[13] < t'$.
Then, when $\alpha$ reads $\X$ into $\x$ at Line~13, the earlier inequalities imply that $\x.\seq \le b$.
From the previous assertion that $\alpha[17] > t''$ and the earlier inequalities, when $\alpha$ reads $\Y$ into $\y$ at Line~17, we have $\y.\seq = b$.
Putting these observations together, we have $\x.\seq \le b = \y.\seq$.
Therefore, when $\alpha$ executes Line~18, the if-condition there evaluates to $\false$; so, $\alpha$ does not execute the CAS at that line, contradicting that $\alpha$ moves.

\vspace{0.1in}
\noindent
\underline{\em Proof of Part (3)}: 
We know from Part (2) that $\alpha[13] > t$.
Thus, we have $t' < \alpha[13] < \alpha[17] < \alpha[18] = t$.
So, when $\alpha$ reads $\X$ into $\x$ at Line~13, $\x.\hndl$ has $\h$ (by Part (1a) of this lemma).
So, when it reads $\x.\hndl.\Val$ into $\vh$ at Line~16, $\vh = \v$ (by Part (1b) of this lemma).
Therefore, when $\alpha$ moves by executing a successful CAS at Line~18, the CAS sets $\Y.\val$ to $\vh = \v$.
Hence, we have Part (3). 
\hfill $\blacksquare$

We define a {\em hitchhiker} as a $\CSC$ operation that does not install and returns at Line~12.

\begin{lemma}\label{hh1-durec}
{\em
If a $\CSC{\h,\s,\v}$ operation $\alpha$ is a hitchhiker, then $\Y.\seq = \s$ at $\alpha[6]$ and $\X.\seq > \Y.\seq$ at some time during the  interval $[\alpha[6], \alpha[10]]$.
}
\end{lemma}

{\em Proof}: The check at Line~6 and the fact that $\alpha$ didn't return at Line~6 imply that $\Y.\seq = \s$ at $\alpha[6]$.
To prove the rest of the lemma, assume to the contrary that $\X.\seq = \Y.\seq$ throughout the interval $I = [\alpha[6], \alpha[10]]$.
Then, it follows from Lemma~\ref{prelim1} that this interval lies entirely between a move and the following install.
Therefore, $\X$  and $\Y$ do not change during this interval $I$ (by Lemma~\ref{prelim1}).
Therefore, when $\alpha$ executes Line~10, $\X.\hndl$ has the same value $\hh$ that it did at Line~8.
So, from $\alpha$'s failed CAS at Line~10, we can infer that $\X.\seq \neq \s$ at $\alpha[8]$.
Since $\X$ does not change during $I$, it follows that $\X.\seq \neq \s$ at $\alpha[6]$.
Then, by Lemma~\ref{all}, $\X.\seq > \Y.\seq$ at $\alpha[6]$, contradicting our assumption.
Hence, we have the lemma.
\hfill $\blacksquare$

The next lemma states that a move occurs during every hitchhiking write.

\begin{lemma}\label{hh-durec}
{\em
If $\alpha$ is a hitchhiker $\CSC$ operation, a move occurs between $\alpha[6]$ and $\alpha[12]$.
}
\end{lemma}

{\em Proof}: By the previous lemma, $\X.\seq > \Y.\seq$ at some time $t$ between $\alpha[6]$ and $\alpha[10]$. Since $\alpha$ executes a $\forward$ operation at Line~11,  after $\alpha[10]$ and before it returns at Line~12, Lemma~\ref{tran1} implies that a move occurs between $t$ and $\alpha[12]$.
Hence, we have the lemma.
\hfill $\blacksquare$

The next definition states how operations are linearized.
A crashed operation is not linearized, unless it is a $\CSC$ operation that crashes after installing.  
Hitchhikers return $\false$ at Line~12, so they are not crashed operations and are linearized.

\begin{definition}[Linearization]\label{wlin1}
{\em

\mbox{ }
\begin{enumerate}
\item
If a $\CSC{\h, \s, \v}$ operation $\alpha$ installs, it is linearized at the first move after $\alpha$'s install.

(Lemma~\ref{wok-durec} guarantees that $\alpha$ is linearized before it completes.)

\item
If a $\CSC{\h, \s, \v}$ operation $\alpha$ is a hitchhiker, it is linearized at the earliest time $t$ such that $t > \alpha[6]$ and a move occurs at $t$.
Furthermore, if $\beta$ is the installing $\CSC$ operation linearized at the same time $t$,
$\alpha$ is linearized {\em after} $\beta$.

{\em Remarks}: Lemma~\ref{hh-durec} guarantees that $\alpha$ is linearized before it completes.
Linearizing a hitchhiker after the installer ensures that the success of the installer's ECSC causes the hitchhikers's ECSC to fail without changing the object's state, thereby eliminating the burden of detecting the hitchhikers' ECSC operation.

\item
If a $\CSC{\h, \s, \v}$ operation $\alpha$ returns at Line~6, it is linearized at $\alpha[6]$.

\item
A $\CLL{\h}$ operation $\alpha$ is linearized at $\alpha[4]$.

\item
A $\CVL{\h, \s}$ operation $\alpha$ is linearized at $\alpha[5]$.
\end{enumerate}
}
\hfill $\blacksquare$
\end{definition}

The value of a DurEC object implemented by the algorithm changes atomically at the linearization points of successful $\CSC$ operations.
The next lemma states that the algorithm maintains the DurEC object's state in $\Y$, and satisfies durable linearizability.

\begin{lemma}[Durable-linearizability of \DurEC \ objects]\label{juice-durec}
{\em 
Let $\O$ be a DurEC object implemented by the algorithm.

\begin{enumerate}
\item
$(\O.\seq, \O.\val) = (\Y.\seq, \Y.\val)$ at all times.
\item
Let $\alpha$ be any $\O.\CSC{\h, \s, \v}$, $\O.\CLL{\h}$, or $\O.\CVL{\h}$ operation, and $t$ be the time at which $\alpha$ is linearized.
Suppose that $\O$'s state is $\sigma$ at $t$ just before $\alpha$'s linearization (in case multiple operations are linearized at $t$), and $\delta(\sigma, \alpha) = (\sigma', r)$, where $\delta$ is the sequential specification of a EC object. Then:

\begin{enumerate}
\item
$\O$'s state changes to $\sigma'$ at time $t$.
\item
If $\alpha$ completes without crashing, it returns $r$.

(Recall that if $\alpha$ crashes and, upon restart, executes $\Recover$, the recover method does not return any response.)
\end{enumerate}

\end{enumerate}
}
\end{lemma}

{\em Proof}: 
We prove the lemma by induction.
The base step follows from the algorithm's initialization that sets $\Y.\val$ to $\O$'s initial value; and since $\O$ is only an abstract object, we notionally set $\O.\seq$'s initial value to $\Y.\seq$'s initial value of 0.
The induction hypothesis is that $i \ge 1$ and the lemma holds up to and including the first $i-1$ linearization times.
Let $\tau$ be the next linearization time and $S$ be the set of operations that are linearized at $\tau$.
For the induction step, we show that the lemma holds after the operations in $S$ take effect at $\tau$ (in their linearization order).
There are many possibilities for what operations are linearized at $\tau$, and in the following, we show the induction step for each possibility.

\begin{itemize}
\item
\underline{Case 1, a move occurs at $\tau$}:
In this case, one installer $\CSC{\h, \s, \v}$ operation $\alpha$ and a (possibly empty) set $S$ of hitchhiker $\CSC$ operations are linearized at $\tau$, with $\alpha$ linearizing before the ones in $S$ (by Definition~\ref{wlin1}). The move at $\tau$ increases $\Y.\seq$ from $\s$ (Lemma~\ref{all}). Hence, by the induction hypothesis, immediately before the move, $\O.\seq = \s$.
So, $\alpha$'s linearization at $\tau$ implies $\alpha$ succeeds, sets $\O.\val$ to $v$, and increases $\O.\seq$. The move at $\tau$  sets $\Y.\val$ to $\v$ (Part 3 of Lemma~\ref{m1}), and $\alpha$ returns $\true$ at Line~12 (because $r$ was set to true when $\alpha$ installed at Line~10). Furthermore, we set $\O.\seq$ to the value of $\Y.\seq$ immediately after the move. Hence, the induction step holds for the installer's linearization.

Turning to hitchhikers, Lemma~\ref{hh1-durec} guarantees that the second argument of each hitchhiker linearized at $\tau$ is $\s$, and $\O.\seq$ has just been increased from $\s$ by the installer's linearization. Consequently, each hitchhiker must fail and return $\false$. This is precisely what happens in the algorithm: at Line~ 10, a hitchhiker sets $r$ to $\false$ (because, by definition, a hitchhiker's install fails), and returns $\false$ at Line~12.
Hence, we have the induction step for this case.

\item
\underline{Case 2, a $\CSC{\h, \s, \v}$ operation $\alpha$, which returns at Line~6, is linearized at $\tau$}:
In this case, given the condition at Line~6, we have $\Y.\seq \neq \s$ at $\alpha[6]$.
Then, by the induction hypothesis, $\O.\seq \neq \s$ at $\alpha[6]$.
So, $\alpha$'s linearization at $\tau$ implies that $\O$'s state does not change and $\O$ returns $\false$ to $\alpha$.
Therefore, the return of $\false$ at Line~6, without changing $\Y$, establishes the induction step for this case.

\item
\underline{Case 3, a $\CLL{\h}$ operation $\alpha$ is linearized at $\tau$}:
In this case, $\tau$ is the time of $\alpha$'s execution of Line~4.
Since $\O.\val = \Y.\val$ at $\tau$ (by the induction hypothesis), $\alpha$'s linearization at $\tau$ implies that $\O$'s state remains unchanged and $\O$'s response is $\Y$.
Therefore, the return of  $\Y$ at Line~4, without changing $\Y$, establishes the induction step for this case.

\item
\underline{Case 4, a $\CVL{\h, \s}$ operation $\alpha$ is linearized at $\tau$}:
In this case, $\tau$ is the time of $\alpha$'s execution of Line~4.
Since $\O.\val = \Y.\val$ at $\tau$ (by the induction hypothesis), $\alpha$'s linearization at $\tau$ implies that $\O$'s state remains unchanged and $\O$'s response is $\Y$.
Therefore, the return of  $\Y$ at Line~4, without changing $\Y$, establishes the induction step for this case.
\end{itemize}

Hence, the induction step is complete and we have the lemma.
\hfill $\blacksquare$

\noindent
{\em Proof}: Line~15 is the only line where the $\detval$ field of any handle might change, and it is clear from the code at this line that, if the $\detval$ field changes, it increases.
\hfill $\blacksquare$

Next we state a key lemma for proving the detectability of  DurEC objects.

\begin{lemma}\label{det1}
{\em
\mbox{ }

\begin{enumerate}
\item
If a $\CSC{\h,\s,\v}$ operation $\alpha$ installs, then the value of $\h.\detval$ increases between  $\alpha$'s invocation and completion.
\item
For any handle $\h$, if $\h.\detval$ is changed at any time $t$ by the execution of Line~15 by some $\forward{\h'}$ method (for some $\h'$), then $\X.\hndl = \h$ and $\h.\pc \in \{13,14,15\}$.
\item
If a $\CSC{\h,\s,\v}$ operation $\alpha$ does not install, then the value of $\h.\detval$ is the same at $\alpha$'s invocation and completion.
\end{enumerate}
}
\end{lemma}

\noindent
{\em Proof}: We prove the lemma using the following statements, which constitute a key invariant that the algorithm maintains:

\begin{enumerate}
\item
If $\pc \in \{14, 15\} \wedge (\x.\hndl.\detval < \x.\seq)$, then $\x = \X$
\item
If $\X.\hndl.\detval < \X.\seq$, then $\X.\hndl.\pc \in \{13, 14, 15\}$
\item
If $\X.\hndl.\pc \in \{13, 14, 15\}$, then $(\X.\hndl.\detval \le \X.\seq) \wedge (\X.\seq > \Y.\seq)$
\item
If $\X.\hndl.\pc \in \{12, 16, 17, 18, 19, 21\}$, then $\X.\hndl.\detval = \X.\seq$
\item
If $\X.\seq > \Y.\seq$, then $\X.\hndl.\detval \le \X.\seq$
\end{enumerate}

We omit the inductive proof of this invariant, but argue how this invariant implies the lemma.

\begin{itemize}
\item
{\em Proof of Part (1) of the lemma}: 
Suppose that a $\CSC{\h,\s,\v}$ operation $\alpha$ installs.
Line~9 ensures that $\sh > \h.\detval$ at $\alpha[7]$.
When $\alpha$ is at Line~10, since $\h.\pc = 10$, Statements (1) and (2) of the above invariant imply that $\h.\detval$ does not change.
Therefore, when $\alpha$ installs at Line~10 by performing a successful CAS on $\X$, $\X.\seq$ becomes $\sh$; so, $\h.\detval < \sh = \X.\seq$ at that point.,,
However, when $\alpha$ completes by returning at Line~12 or Line~21, Statement (4) of the invariant implies that $\h.\detval = \X.seq$; furthermore, since $\X.\seq$ never decreases, $\X.\seq \ge \sh$ at that point.
Thus, $\h.\detval < \sh$ at $\alpha[10]$, and $\h.\detval <\ge \sh$ at $\alpha$'s completion.
Hence, we have Part (1) of the lemma.

\item
{\em Proof of Part (2) of the lemma}:
Statements (1) and (2) of the above invariant imply that $\h.\detval$ is not changed by Line~15 anytime unless $\X.\hndl = \h$ and $\h.\pc \in \{13,14,15\}$.
\hfill $\blacksquare$

\item
{\em Proof of Part (3) of the lemma}:
Suppose that a $\CSC{\h,\s,\v}$ operation $\alpha$ does not install.
Then, during $\alpha$'s execution, it is never the case that $\X.\hndl = \h$ and $\pc \in \{13,14,15\}$.
This, together with Part (2) of the lemma, implies Part (3).
\hfill $\blacksquare$
\end{itemize}

\begin{lemma}[Detectability of \DurEC \ objects]\label{det2}
{\em
Let $\alpha$ be any operation executed on a DurEC object $\O$ by a handle $\h$.
Suppose that $(d_1, r_1)$ and $(d_2, r_2)$ are the values that $\Detect{\h}$ would return, if executed immediately before $\alpha$ is invoked and immediately after $\alpha$ completes, respectively.
Then:
\begin{enumerate}
    \item
    If $\alpha$ is not an installing ECSC, it is safe to repeat and $d_2 = d_1$.
    \item
    If $\alpha$ is an installing ECSC, then $d_2 > d_1$ and $r_2 = \true$.
\end{enumerate}
}
\end{lemma}

{\em Proof}:  For Part (1), hitchhiking ECSC operations, ECSC operations that return at Line~6, ECLL and ECVL operations do not change $\O$'s state, and hence are safe to repeat. Furthermore, Parts (2) and (3) of Lemma~\ref{det2}, together with the $\Detect$ method's code at Line~22,  imply that $d_1 = d_2$ for any of these operations.
Hence, we have Part (1) of the lemma.

If $\alpha$ is an installing $\CSC$ operation, Part (3) of Lemma~\ref{det2}, together with the $\Detect$ method's code at Line~22,  imply that $d_2 > d_1$ and $r_2 = \true$.
Hence, we have Part (2) of the lemma.
\hfill $\blacksquare$

\durec*

{\em Proof}: The first property follows from Lemmas \ref{juice-durec} and \ref{det2}.
The second property is clear from an inspection of the algorithm.
The third property follows from the observation that the algorithm makes no assumption about the the number or the names of processes that participate in the algorithm: any process can begin participating by creating a handle for itself, and accessing any existing DurEC objects, or creating new ones by calling the constructor.
For the fourth property, we note that
each DurEC handle $\h$ needs space for two fields ($\detval$ and $\Val$), and
each DurEC object $\O$ needs space for two variables ($\X$ and $\Y$).
Therefore, if $n$ DurEC handles and $m$ DurEC objects are created in a run, the space required is $O(m+n)$.
\hfill $\blacksquare$

\subsection{Proof of correctness of {\sc DurECW}}
\label{DurECWProof}

If a $\Write$ or a $\CSC$ method crashes while executing an operation on $\W$ or $\Z$, upon restart, Lines 21 to 24 of $\Recover$ ensure that $\W.\Recover$ or $\Z.\Recover$ is executed before any other operation is executed on $\W$ or $\Z$.  Consequently, the durable objects $\W$ and $\Z$ behave like atomic EC objects, and are treated as such in the rest of the proof. 


We introduce the following terms. 

An {\em install} is a successful ECSC operation on $\W$, executed by a $\Write$ operation at its Line~14.
A $\Write$ operation {\em installs a value $\v$ in $\W$ at time $t$} if it executes a successful ECSC on $\W$ at Line~14 at time $t$ and sets $\W.\val$ to $\v$.

A {\em move} is a successful ECSC operation on $\Z$, executed by a $\transfer$ operation at its Line~20.
A $\transfer$ operation {\em moves a value $\v$ to $\Z$ at time $t$} if it executes a successful ECSC on $\Z$ at Line~20 at time $t$ and sets $\Z.\val$ to $\v$.

An {\em imprint} is a successful ECSC operation on $\Z$, executed by a $\CSC$ operation at its Line~10.
A $\CSC$ operation {\em imprints a value $\v$ in $\Z$ at time $t$} if it executes a successful ECSC on $\Z$ at Line~10 at time $t$ and sets $\Z.\val$ to $\v$.

A $\Write$ operation is a {\em hitchhiker} if it does not install and returns at Line~17.

To refer to the times when the algorithm performs various actions, we adopt the following notation.
If $\alpha$ is an execution of a $\Write$, $\CSC$, or $\transfer$ and $\ell$ is a line number in the algorithm, $\alpha[\ell]$ denotes the time at which $\alpha$ executes Line $\ell$.

We denote an open interval from time $t_1$ to $t_2$ by $(t_1,t_2)$, the closed interval by $[t_1,t_2]$, and the semi-closed interval that includes $t_1$ and not $t_2$ by $[t_1,t_2)$.

\begin{lemma}\label{inst1-durecw}
{\em
\mbox{ }

\begin{enumerate}
\item
If a $\Write{\h,\v}$ operation $\alpha$ installs, then
$\W$'s state remains the same in the interval $[\alpha[12], \alpha[14])$, and
at time $\alpha[14]$, $\W.\bit$ flips and $\W.\val$ becomes $v$.
\item
If a $\transfer{\h}$ operation $\alpha$ moves,
then $\Z$'s state remains the same in the interval $[\alpha[18], \alpha[20])$, and at time $\alpha[20]$, $\Z.\bit$ flips.
\item
If a $\CSC{\h, \s, \v}$ operation $\alpha$ imprints,
then $\Z$'s state remains the same in the interval $[\alpha[7], \alpha[10])$.
Furthermore, at time $\alpha[10]$, $\Z.\val$ is set to $\v$, $\Z.\seq$ increases from $\s$, and $\Z.\bit$ is unchanged.
\end{enumerate}
}
\end{lemma}

{\em Proof}: In Part (1), at time $\alpha[14]$, since the ECSC succeeds, it must be $\W.\seq = \w.\seq$. 
Since $\W.\seq$ was $\w.\seq$ back at $\alpha[12]$, it follows that $\W.\seq = \w.\seq$ all through the interval $[\alpha[12], \alpha[14])$, which implies that $\W$'s state remains the same all through that interval.
Since $\w = \W$ when the ECSC succeeds, the ECSC sets $\W.\val$ to $v$ and flips $\W.\bit$. Hence, we have Part (1).
Parts (2) and (3) are proved analogously. 
\hfill $\blacksquare$

The next lemma states that $\W$ changes and $\W.\bit$ flips precisely at times of installs, $\Z$ changes precisely at times of moves or imprints, and $\Z.\bit$ flips only at moves, and not at imprints.

\begin{lemma}\label{obs-durecw}
{\em
\mbox{ }
\begin{enumerate}
\item
At any time $t$, $\W$'s state changes or $\W.\bit$ flips at $t$ if and only if an install occurs at $t$.
\item
At any time $t$, $\Z$'s state changes at $t$ if only if a move or an imprint occurs at $t$.
\item
At any time $t$, $\Z.\bit$ flips at $t$ if and only if a move occurs at $t$.
\end{enumerate}
}
\end{lemma}

{\em Proof}: 
A successful ECSC on $\W$ at Line~14 changes $\W$'s state and flips $\W.\bit$, and no other action in the algorithm changes $\W$.
A successful ECSC on $\Z$ at Line~10 or Line~20 changes $\Z$'s state and no other action in the algorithm changes $\Z$; moreover, only a successful ECSC on $\Z$ at Line~20 flips $\Z.\bit$.
Hence, we have the lemma.
\hfill $\blacksquare$

\begin{lemma}\label{obs2-durecw}
{\em
\mbox{ }

\begin{enumerate}
\item
Suppose that $\alpha'$ and $\alpha$ are two different $\Write$ operations that install, and $\alpha'$ installs before $\alpha$, i.e., $\alpha'[14] < \alpha[14]$. Then, $\alpha'[14] < \alpha[12]$.
\item
Suppose that $\alpha$ and $\alpha'$ are two different $\transfer$ operations that move, and $\alpha'$ moves before $\alpha$, i.e., $\alpha'[20] < \alpha[20]$. Then, $\alpha'[20] < \alpha[18]$.
\item
Suppose that $\alpha'$ and $\alpha$ are two different $\CSC$ operations that imprint, and $\alpha'$ imprints at $\alpha'[10]$ before $\alpha$ imprints at $\alpha[10]$. Then, $\alpha'[10] < \alpha[7]$.
\end{enumerate}
}
\end{lemma}

{\em Proof}: For a proof of Part (1), assume to the contrary that $\alpha[12] < \alpha'[14] < \alpha[14]$.
At $\alpha'[14]$, since the ECSC on $\W$ succeeds, $\W.\seq$ changes,
contradicting Part (1) of Lemma~\ref{inst1-durecw} that $\W$'s state remains the same in the interval $[\alpha[12], \alpha[14])$.
The other two parts of the lemma are proved analogously.
\hfill $\blacksquare$

\begin{lemma}\label{ifirst-durecw}
{\em
An install occurs before any move occurs.
}
\end{lemma}

{\em Proof}: Assume to the contrary that a move occurs at time $t$ and no install occurs before $t$.
Let $\alpha$ be the $\transfer$ operation that moves at $t$.
Since no installs or moves occur before $t$, Lemma~\ref{obs-durecw} implies that $\W.\bit$ and $\Z.\bit$ are never flipped before $t$.
So these bits still have their initial value of 0 when $\alpha$ executes ECLL at lines 18 and 19 to read the states of $\Z$ and $\W$ into $z$ and $w$, respectively.
Thus, $\z.\bit = \w.\bit = 0$ when $\alpha$ executes Line~20.
Therefore, the if-condition at Line~20 prevents $\alpha$ from executing the ECSC operation, contradicting that $\alpha$ moves.
\hfill $\blacksquare$

The next lemma states that installs and moves alternate.

\begin{lemma}\label{rhythm-durecw}
{\em
Between any two moves, there is an install; and between any two installs, there is a move.
}
\end{lemma}

{\em Proof}: We consider two cases---that installs occur consecutively with no intervening moves or that moves occur consecutively with no intervening installs---and derive a contradiction in each case.

Suppose that the lemma is violated for the first time when an install occurs at time $t$, following an install at an earlier time $t' < t$ and with no moves in the interval from $t'$ to $t$.
Since the lemma was not violated before $t$, Lemma~\ref{obs-durecw} implies that, counting $\W$'s flip at $t'$, $\W$ is flipped one more time than $\Z$ by time $t'$, and no further flips happen to either bit in the open interval from $t'$ to $t$.
It follows that, in the open interval from $t'$ to $t$, the values of $\W.\bit$ and $\Z.\bit$ do not change and, since $\W.\bit$ and $\Z.\bit$ are both initially 0, $\W.\bit \neq \Z.\bit$.
Let $b$ and $1-b$ be the stable values of $\W.\bit$ and $\Z.\bit$, respectively, in the interval $(t',t)$.
Let $\alpha'$ and $\alpha$ denote the $\Write$ operations that install at times $t'$ and $t$, respectively; thus, $t' = \alpha'[14]$ and $t = \alpha[14]$.
From Part (1) of Lemma~\ref{obs2-durecw}, we know that $\alpha'[14] < \alpha[12] < \alpha[14]$.
Then, since $\W.\bit$ and $\Z.\bit$ remain unchanged at $b$ and $1-b$ during the open interval $(\alpha'[14], \alpha[14])$, when $\alpha$ executes ECLL at Lines 12 and 13 to read $\W$'s state and $\Z$'s state into $w$ and $z$, respectively, $\w.\bit$ is set to $b$ and $\z.\bit$ is set to $1-b$.
Therefore, when $\alpha$ executes Line~14, the if-condition there evaluates to {\em false}, preventing $\alpha$ from executing the ECSC operation, which contradicts that $\alpha$ installs.
By an analogous argument, a contradiction arises in the second case also.
Hence, we have the lemma.
\hfill $\blacksquare$

\begin{corollary}\label{seesaw-durecw}
{\em
\mbox{ }

\begin{enumerate}
\item
$\W.\bit$ and $\Z.\bit$ do not change and have the same value in the open interval up to the first install and in each open interval from a move to the first install that follows that move.
\item
$\W.\bit$ and $\Z.\bit$ do not change and have different values in each open interval from an install to the first move that follows that install.
\end{enumerate}
}
\end{corollary}

{\em Proof}: Follows from Lemmas \ref{obs-durecw} and \ref{rhythm-durecw}, and the intialization of $\W.\bit$ and $\Z.\bit$ to 0.
\hfill $\blacksquare$

\begin{lemma}\label{snap-durecw}
{\em
\mbox{ }

\begin{enumerate}
\item
If a $\Write$ operation $\alpha$ installs, then $\W.\bit = \Z.\bit$ at time $\alpha[13]$.
\item
If a $\transfer$ operation $\alpha$ moves, then $\W.\bit \neq \Z.\bit$ at time $\alpha[19]$.
\end{enumerate}
}
\end{lemma}

{\em Proof}:
Since $\alpha$ is successful, it follows from Lemma~\ref{inst1-durecw} that $\W$'s state does not change in the open interval from $\alpha[12]$ to $\alpha[14]$.
Therefore, $\w$, which is $\W$'s state returned by $\alpha$'s ECLL at Line~12, continues to be $\W$'s state at $\alpha[14]$.
The ECLL on $\Z$ at Line~13 returns $\Z$'s state at $\alpha[10]$ into $\z$.
Thus, the values of $\W.\bit$ and $\Z.\bit$ at time $\alpha[10]$ are $\w.\bit$ and $\z.\bit$, respectively.
Since $\alpha$ installs (i.e., executes a successful ECSC at Line~14), the if-condition at Line~14 must evaluate to {\em true}, which implies that $\z.\bit = \w.\bit$, which from the above implies that $\W.\bit = \Z.\bit$ at time $\alpha[13]$.
Hence, we have Part (1).
Part (2) is proved analogously.
Hence, we have the lemma.
\hfill $\blacksquare$

\begin{lemma}\label{wtozhelp-durecw}
{\em
\mbox{ }

\begin{enumerate}
\item
If an install occurs at time $t$ and the first move after $t$ is executed by a $\transfer$ operation $\alpha$, then $\alpha[19] > t$.
\item
If a move occurs at time $t$, and the first install after $t$ is executed by a $\Write$ operation $\alpha$, then $\alpha[13] > t$.
\end{enumerate}
}
\end{lemma}

{\em Proof}: For a proof of Part (1), assume for a contradiction that $\alpha[19] < t < \alpha[20]$.
Part (2) of Lemma~\ref{snap-durecw} implies that $\W.\bit \neq \Z.\bit$ at time $\alpha[19]$.
Therefore, the instant $\alpha[19]$ falls in a time interval where $\W.\bit \neq \Z.\bit$.
If the install at $t$ is the first install, let $t' = 0$; otherwise, let $t'$ be the latest time before $t$ when a move occurs.
Corollary~\ref{seesaw-durecw} implies that $\W.\bit = \Z.\bit$ during the open interval from $t'$ to $t$, and $\W.\bit \neq \Z.\bit$ during the open interval from $t$ to $\alpha[20]$.
Then, since $\alpha[19] < t$ and $\alpha[19]$ falls in a time interval where $\W.\bit \neq \Z.\bit$,
it must be that $\alpha[19] < t'$.
It follows that $t' > 0$ and the install at $t$ is not the first install.
From the above, we have $\alpha[19] < t' < t < \alpha[20]$.
Since a move occurs at $t'$, there is a successful ECSC on $\Z$ at $t'$.
Therefore, $\Z$'s state changes at $t'$, which implies that
$\Z$'s state changes between $\alpha[19]$ and $\alpha[20]$, contradicting Part (2) of Lemma~\ref{inst1-durecw}, which states that $\Z$'s state does not change in the interval from $\alpha[18]$ to $\alpha[20]$.
Hence, we have Part (1).
Part (2) is proved analogously.
Hence, we have the lemma.
\hfill $\blacksquare$

The next lemma states that if two $\transfer$ operations are executed after installing a value $v$ in $\W$, $v$ is sure to move to $\Z$ by the time the second $\transfer$ operation completes.

\begin{lemma}\label{tran-durecw}
{\em
If $\W.\bit \neq \Z.\bit$ at time $t$,
a $\transfer$ operation $\alpha_1$ is started after $t$, and another $\transfer$ operation $\alpha_2$ is started after $\alpha_1$ completes, then a move occurs  after $t$ and at or before $\alpha_2$'s completion time of $\alpha_2[20]$.
}
\end{lemma}

{\em Proof}: Assume to the contrary that no move occurs between $t$ and $\alpha_2[20]$.
Then, Corollary~\ref{seesaw-durecw} implies that, in the interval from $t$ to $\alpha_2[20]$, $\W.\bit$ and $\Z.\bit$ do not change and $\W.\bit \neq \Z.\bit$.
Let $b \in \{0,1\}$ and $1-b$ be $\W.\bit$'s and $\Z.\bit$'s values, respectively, in this interval.
Therefore, in both $\alpha_1$ and $\alpha_2$, the ECLL operations on $\Z$ and $\W$ at Lines 18 and 19 return $z$ and $w$ such that $z.\bit = 1-b$ and $w.\bit = b$;
so, the if-condition at Line~20 evaluates to {\em true} and the ECSC operation on $\Z$ is attempted.
However, by our assumption that no move occurs by time $\alpha_2[20]$, the ECSC on $\Z$ at Line~20 is unsuccessful in both $\alpha_1$ and $\alpha_2$.

The failure of $\alpha_1$'s ECSC at Line~20 must be because $\Z$'s state changed between $\alpha_1$'s ECLL at Line~18 and its subsequent ECSC at Line~20.
Therefore, by Lemma~\ref{obs-durecw}, an imprint or a move occurs between $\alpha_1[18]$ and $\alpha_1[20]$.
It must be an imprint because a move is ruled out by our assumption that no move occurs by time $\alpha_2[20]$.
Let $\beta_1$ be the $\CSC$ operation that imprints between $\alpha_1[18]$ and $\alpha_1[20]$.
Arguing similarly, there is a $\CSC$ operation $\beta_2$ that imprints between $\alpha_2[18]$ and $\alpha_2[20]$.
Let $i_1 \in \{1,2\}$ and $i_2 \in \{1,2\}$ be the iterations (of the for-loop at Line~6*) during which $\beta_1$ and $\beta_2$, respectively, execute a successful ECSC on $\Z$.
Part (3) of Lemma~\ref{inst1-durecw} implies that $\Z$'s state is unchanged in the open interval from $\beta_2[7,i_2]$ to $\beta_2[10,i_2]$.

Since $\beta_1$ imprints between $\alpha_1[18]$ and $\alpha_1[20]$, we have $\alpha_1[18] < \beta_1[10,i_1] < \alpha_1[20]$.
Similarly, we have $\alpha_2[18] < \beta_2[10,i_2] < \alpha_2[20]$.
Furthermore, Part (3) of Lemma~\ref{obs2-durecw} implies that $\beta_1[10,i_1] < \beta_2[7,i_2]$.
Putting these together, we have $t < \alpha_1[18] < \beta_1[10,i_1] < \beta_2[7,i_2] < \beta_2[10,i_2] < \alpha_2[20]$.
In particular, $t < \beta_2[7,i_2] < \beta_2[10,i_2] < \alpha_2[20]$.
Furthermore, Part (3) of Lemma~\ref{inst1-durecw} implies that $\Z$'s state is unchanged in the open interval from $\beta_2[7,i_2]$ to $\beta_2[10,i_2]$; in particular, $\Z$'s state is constant throughout the time that $\beta_2$ calls the $\transfer$ method from its Line$[9,i_2]$ and executes it.
The last two observations imply that, during this execution of $\transfer$,
$\beta_2$'s ECLL on $\Z$ at Line~18 returns a $\zh$ such that $\zh.\bit = 1-b$, its ECLL on $\W$ at Line~19 returns a $\wh$ such that $\wh.\bit = b$, and $\Z$'s state is still $\zh$ when $\beta_2$ executes Line~20.
It follows that the if-condition at Line~20 evaluates to {\em true}, so $\beta_2$ executes the ECSC operation; and this ECSC succeeds because $\Z$'s state is still $\zh$.
This successful ECSC means that $\beta_2$ moves, contradicting our assumption that no move occurs between $t$ and $\alpha_2[20]$.
\hfill $\blacksquare$

The next lemma states that if a $\Write$ operation $\alpha$ installs a value $v$ in $\W$, that value is moved to $\Z$ before $\alpha$ completes.

\begin{lemma}\label{wok-durecw}
{\em
If a $\Write{\h,\v}$ operation $\alpha$ installs $\v$ at time $t$, the first move after $t$ occurs before $\alpha$ completes and it moves $\v$ to $\Z$.
}
\end{lemma}

{\em Proof}: After $\alpha$ installs at time $t$ (by executing a successful ECSC on $\W$ at Line~14), $\alpha$ executes the $\transfer$ method twice, at Lines 15 and 16.
If the process $p$ executing $\alpha$ crashes before executing the $\transfer$ method twice, upon restart, $p$ executes the {\sc recover} method where, after recovering from any partially executed operations on $\W$ and $\Z$ at Lines 21 to 24, $\transfer$ is executed twice, at Lines 25 and 26.
Thus, after $\alpha$ installs at time $t$, regardless of crashes, $\alpha$ executes $\transfer$ at least twice, before completing.
It follows from Lemma~\ref{tran-durecw} that a move occurs before $\alpha$ is completed.
Hence, we have the first part of the lemma.

To complete the lemma, we prove that the first move after $t$ moves $\v$ to $\Z$.
Let $\alpha'$ be the $\transfer$ operation that executes the first move after $t$.
$\W.\val$ is set to $v$ at time $t$ (by Part (1) of Lemma~\ref{inst1-durecw}), and it remains unchanged until the next install (by Part (1) of Lemma~\ref{obs-durecw}).
Since a move must occur between any two installs (by Lemma~\ref{rhythm-durecw}), it follows that $\W.\val = \v$ during the interval from $t$ to $\alpha'[20]$.
Furthermore, $\alpha'[19] > t$ (by Lemma~\ref{wtozhelp-durecw}).
It follows from the above that, when $\alpha'$ reads $\W$ into $\wh$ at Line~19, $\wh.\val = \v$.
Therefore, when $\alpha'$ moves by executing a successful ECSC on $\Z$ at Line~20, the passing of $\wh.\val$ as the third argument of that ECSC ensures that $\Z.\val$ is set to $\v$.
Thus, the first move after $t$ moves $v$ to $\Z$.
Hence, we have the lemma.
\hfill $\blacksquare$

Recall that a hitchhiker is a $\Write$ operation that does not install and returns at Line~17.

\begin{lemma}\label{hh1-durecw}
{\em
If $\alpha$ is a hitchhiker $\Write$ operation, $\W.\bit \neq \Z.\bit$ at some time during the semi-closed interval $(\alpha[12], \alpha[14]]$.
}
\end{lemma}

{\em Proof}: Assume to the contrary that $\W.\bit = \Z.\bit$ throughout the interval $(\alpha[12], \alpha[14]]$.
Then, Corollary~\ref{seesaw-durecw} implies that in this interval $\W$'s state and $\Z.\bit$ do not change their values.
It follows that, when $\alpha$ executes an ECLL on $\W$ and $\Z$ at Lines 12 and 13, the return values $\w$ and $\z$ are such that $\w.\bit = \z.\bit$ and $\W$'s state continues to be $\w$ at $\alpha$'s Line~14.
Since $\alpha$ is a hitchhiker, it does not return at Line~13*;
furthermore, at Line~14, since $\w.\bit = \z.\bit$, the if-condition evaluates to {\em true}, and since $\W = \w$, the ECSC succeeds.
Thus, $\alpha$ installs, contradicting that $\alpha$ is a hitchhiker.
\hfill $\blacksquare$

The next lemma states that a move occurs during every hitchhiking write.

\begin{lemma}\label{hh-durecw}
{\em
If $\alpha$ is a hitchhiker $\Write$ operation, a move occurs between $\alpha[12]$ and $\alpha[17]$.
}
\end{lemma}

{\em Proof}: By the previous lemma, $\W.\bit \neq \Z.\bit$ at some time $t$ between $\alpha[12]$ and $\alpha[14]$. Since $\alpha$ executes two $\transfer$ operations after $\alpha[14]$ and before it returns at Line~17, Lemma~\ref{tran-durecw} implies that a move occurs between $t$ and $\alpha[17]$.
Hence, we have the lemma.
\hfill $\blacksquare$

The next definition states how a $\Write{\h,\v}$ operation is linearized, based on whether it installs or hitchhikes.
A crashed $\Write$ operation is not linearized, unless it unless it crashes after installing.  
Hitchhikers return at Line~17, so they are not crashed operations and are linearized.

\begin{definition}[Linearization of Write]\label{wlin-durecw}
{\em
Let $\alpha$ be a $\Write{\h,\v}$ operation.

\begin{enumerate}
\item
If $\alpha$ installs, it is linearized at the first move after $\alpha$'s install.

(Lemma~\ref{wok-durecw} guarantees that $\alpha$ is linearized before it completes.)

\item
If $\alpha$ is a hitchhiker, let $t$ be the earliest time when a move occurs in $\alpha$'s interval, and $\beta$ be the unique installer linearized at $t$.
(Lemma~\ref{hh-durecw} guarantees that $t$ is well defined, and Lemma~\ref{rhythm-durecw} guarantees that a unique installer is linearized at $t$.)
Then, $\alpha$ is linearized at $t$ (along with $\beta$), and is ordered {\em before} $\beta$.
(This ordering ensures that the hitchhikers are overwritten instantly by the installer, thereby eliminating the burden of detecting the hitchhikers' write operations.)
\end{enumerate}
}
\hfill $\blacksquare$
\end{definition}

Next we state how $\CSC$, $\CLL$, and $\CVL$ operations are linearized.
We do not linearize a crashed $\CSC$ operation, unless it crashes after imprinting.

\begin{definition}[Linearization of $\CSC$, $\CLL$, and $\CVL$]\label{clin-durecw}
{\em
\mbox{ }

\begin{enumerate}
\item
If a $\CSC{\h, \s, \v}$ operation $\alpha$ imprints, it is linearized at $\alpha[10]$, the time at which $\alpha$ imprints.
\item
If a $\CSC{\h, \s, \v}$ operation $\alpha$ returns $\false$ at Line~11, it is linearized at $\alpha[10]$, the time when $\alpha$ performs an unsuccessful ECSC on $\Z$.
\item
If a $\CSC{\h, \s, \v}$ operation $\alpha$ returns at Line~8, it is linearized at $\alpha[7]$.
\item
If a $\CLL{\h}$ operation $\alpha$ returns, it is linearized at $\alpha[4]$.
\item
If a $\CVL{\h, \s}$ operation $\alpha$ returns, it is linearized at $\alpha[6]$.
\end{enumerate}
}
\hfill $\blacksquare$
\end{definition}

The value of a DurECW object implemented by the algorithm changes atomically at points where $\Write$ and $\CSC$ operations are linearized.
The next lemma states that the algorithm maintains the DurECW object's state in $\Z$, and satisfies durable linearizability.

\begin{lemma}[Durable-linearizability of \DurECW \ objects]\label{juice-durecw}
{\em 
Let $\O$ be a DurECW object implemented by the algorithm.

\begin{enumerate}
\item
$\O.\val = \Z.\val$ and $\O.\seq = \Z.\seq$ at all times.
\item
Let $\alpha$ be any $\O.\CSC{\h, \s, \v}$, $\O.\Write{\h, \v}$, $\O.\CLL{\h}$, or $\O.\CVL{\h}$ operation, and $t$ be the time at which $\alpha$ is linearized.
Suppose that $\O$'s state is $\sigma$ at $t$ just before $\alpha$'s linearization (in case multiple operations are linearized at $t$), and $\delta(\sigma, \alpha) = (\sigma', r)$, where $\delta$ is the sequential specification of a ECW object. Then:

\begin{enumerate}
\item
$\O$'s state changes to $\sigma'$ at time $t$.
\item
If $\alpha$ completes without crashing, it returns $r$.

(Recall that if $\alpha$ crashes and, upon restart, executes $\Recover$, the recover method does not return any response.)
\end{enumerate}

\end{enumerate}
}
\end{lemma}

{\em Proof}: 
We prove the lemma by induction.
The base step follows from the algorithm's initialization that sets $\Z.\val$ to $\O$'s initial value; and since $\O$ is only an abstract object, we notionally set $\O.\seq$'s initial value to $\Z.\seq$'s initial value of 0..
The induction hypothesis is that $i \ge 1$ and the lemma holds up to and including the first $i-1$ linearization times.
Let $\tau$ be the next linearization time and $S$ be the set of operations that are linearized at $\tau$.
For the induction step, we show that the lemma holds even after the operations in $S$ take effect at $\tau$ (in their linearization order).
There are many possibilities for what operations are linearized at $\tau$, and we show the induction step for each possibility.

\begin{itemize}
\item
\underline{Case 1, a move occurs at $\tau$}:
In this case, a set $S$ of hitchhiker $\Write$ operations and one installing $\Write{\h,\v}$ operation $\alpha$ are linearized, all at $\tau$, with $\alpha$ linearizing after the ones in $S$ (by Definition~\ref{wlin-durecw}). As a result, all hitchhiker writes are overwritten by $\alpha$, and $\O.\val$ becomes $v$ after all of the operations are linearized. 
Furthermore, the move at $\tau$ sets $\Z.\val$ to $\v$ (by Lemma~\ref{wok-durecw}), thereby ensuring that $\O.\val = \Z.\val$ immediately after $\tau$.


\item
\underline{Case 2, an imprint occurs at $\tau$}:
In this case, the $\CSC{\h, \s, \v}$ operation $\alpha$ that imprints at $\tau$ is the only operation linearized at $\tau$.
Then, $\Z.\val$ is set to $\v$ and $\Z.seq$ is increased from $\s$ at $\tau$ (by Part (3) of Lemma~\ref{inst1-durecw}).
By the induction hypothesis, immediately before $\alpha$'s linearization at $\tau$, $\O.\val = \Z.\val$ and $\O.\seq = \Z.\seq = \s$, and $\alpha$'s linearization sets $\O.\val$ to $\v$, and increases $\O.\seq$ from $\s$. 
We stipulate that this notional increase be such that $\O.\seq$ continues to be the same as $\Z.\seq$ after linearization.
Furthermore, if $\alpha$ does not crash after imprinting (i.e., after the successful ECSC at Line~10), it returns the correct response of {\em true} at Line~11.

\item
\underline{Case 3, a $\CSC{\h, \s, \v}$ operation $\alpha$, which returns $\false$ at Line~11, is linearized at $\tau$}:
In this case, by Part (2) of Definition~\ref{clin-durecw}, $\alpha$ performs an unsuccessful ECSC on $\Z$ at $\tau$, which implies that $\Z.\seq \neq \s$ at $\tau$.
By the induction hypothesis, $\O.\seq \neq \s$ at $\tau$.
Therefore, $\alpha$'s linearization at $\tau$ does not change $\O$'s state, and $\O$ returns $\false$ to $\alpha$.
Since $\Z$ is not changed at $\tau$ and $\alpha$ returns $\false$ at Line~11, the induction step holds for this case.

\item
\underline{Case 4, a $\CSC{\h, \s, \v}$ operation $\alpha$, which returns at Line~8, is linearized at $\tau$}:
In this case, by Definition~\ref{clin-durecw}, $\tau$ is the time of $\alpha$'s execution of Line~7.
It follows from the code at Lines 7 and 8 that at $\tau$, $\Z.\seq \neq \s$.
By the induction hypothesis, $\O.\seq \neq \s$ at $\tau$.
Therefore, $\alpha$'s linearization at $\tau$ does not change $\O$'s state, and $\O$ returns $\false$ to $\alpha$.
Since $\Z$ is not changed at $\tau$ and $\alpha$ returns $\false$ at Line~8, the induction step holds for this case.

\item
\underline{Case 5, a $\CLL{\h}$ operation $\alpha$ is linearized at $\tau$}:
In this case, $\tau$ is the time of $\alpha$'s execution of Line~4.
Since $\O.\val = \Z.\val$ at $\tau$ (by the induction hypothesis), $\alpha$'s linearization at $\tau$ implies that $\O.\val$ remains unchanged, and $\O$'s response is $\Z.\val$ at $\tau$, which is $z.\val$.
This justifies $\alpha$ returning $z.\val$ at Line~5, without changing $\Z$.

\item
\underline{Case 6, a $\CVL{\h, \s}$ operation $\alpha$ is linearized at $\tau$}:
In this case, $\tau$ is the time of $\alpha$'s execution of Line~6.
Since $\O.\seq = \Z.\seq$ at $\tau$ (by the induction hypothesis), $\alpha$'s linearization at $\tau$ implies that $\O$'s response should be the same the $\Z.\CSC$'s response at $\tau$, as implemented by the code at Line~6.
\end{itemize}

Hence, the induction step is complete and we have the lemma.
\hfill $\blacksquare$

Next we prove that a DurECW object $\O$ implemented by the algorithm is detectable.
The key to achieving detectability lies in limiting the use of the handle $\h.\Critical$ to Lines 14 and 10, where install and imprint are attempted.
In particular, when a move is attempted at Line~20, the algorithm employs $\h.\Casual$, and not $\h.\Critical$.
This discrimination ensures that a visible operation---an installing write or an imprinting CAS that affect $\O$'s state in a manner that future operations might witness---increases the ``detector value'' associated with the handle $\h.\Critical$, while safe-to-repeat operations---hitchhiking writes and $\CSC$ operations that return without affecting $\O$'s state (like the ones that return at Line~8 or return $\false$ at Line~11), or operations that crash without affecting $\O$'s state---do not increase the detector value associated with $\h.\Critical$.

\begin{lemma}[Detectability of \DurECW \ objects]\label{det-durecw}
{\em
Let $\alpha$ be any operation executed on a DurECW object $\O$ by a handle $\h$.
Suppose that $(d_1, r_1)$ and $(d_2, r_2)$ are the values that $\Detect{\h}$ would return, if executed immediately before $\alpha$ is invoked and immediately after $\alpha$ completes, respectively.
Then:
\begin{enumerate}
    \item
    If $\alpha$ is neither an installing write nor an imprinting ECSC, it is safe to repeat and $d_2 = d_1$.
    \item
    If $\alpha$ is an installing write or an imprinting ECSC, then $d_2 > d_1$ and $r_2 = \true$.
\end{enumerate}
}
\end{lemma}

{\em Proof}:  
For Part (1), a hitchhiking write operation changes $\O$'s state, but the change is rendered invisible to all future operations because it is instantly overwritten by an installing writer.
$\CSC$ operations that return at Line~8 and $\CSC$ operations that return $\false$ at Line~11, as well as $\CLL$ and $\CVL$ operations, do not change $\O$'s state.
Hence, all of these operations are safe to repeat.
Furthermore, none of these operations perform a successful ECSC on $\W$ or $\Z$ using the handle $\h.\Critical$
(they might perform a successful ECSC at Line~20, but the handle used there is not $\h.\Critical$).
Therefore, DurEC's detectability property implies that calls to $\Detect{\h.\Critical}$ before and after $\alpha$ would return $(d_1, -)$ and $(d_2, -)$ such that $d_1 = d_2$.
So, the call to $\Detect{\h.\Critical}$ at Line~27 establishes the second half of Part (1) of the lemma.

For Part (2), suppose that $\alpha$ is an installing write or an imprinting $\CSC$.
In both cases, $\alpha$ performs a successful ECSC operation $op$ on one of $\W$ or $\Z$, using the handle $\h.\Critical$ (this happens at Line~14 if $\alpha$ is an installing write, and at Line~10 if it is an imprinting $\CSC$).
Then, by the detectability of DurEC objects $\W$ and $\Z$, it follows that 
$\Detect{\h.\Critical}$, if executed before $op$ and after $op$ would return $d_1$ and $d_2$ such that $d_2 > d_1$.
This fact, together with how $\Detect{\h}$ is implemented by Line~27, implies Part (2) of the lemma.
\hfill $\blacksquare$

\durlindurecw*

{\em Proof}: The first property was proved in Lemmas \ref{juice-durecw} and \ref{det-durecw}.
The second property is clear from an inspection of the algorithm.
The third property follows from the observation that the algorithm makes no assumption about the the number or the names of processes that participate in the algorithm: any process can begin participating by creating a handle for itself, and accessing any existing DurECW objects, or creating new ones by calling the constructor.
For the fourth property, we note that
each DurECW handle $\h$ needs space for two DurEC handles ($\h.\Critical$ and $\h.\Casual$), and
each DurECW object $\O$ needs space for two DurEC objects ($\W$ and $\Z$).
Since each DurEC handle and DurEC object requires only $O(1)$ space, if $n$ DurECW handles and $m$ DurECW objects are created in a run, the space required is $O(m+n)$.
\hfill $\blacksquare$

\subsection{Proof of correctness of {\sc DuraCAS}}
\label{DuraCasProof}

If a $\Write$ or a $\CAS$ method crashes while executing an operation on $\W$ or $\Z$, upon restart, Lines 21 to 24 of $\Recover$ ensure that $\W.\Recover$ or $\Z.\Recover$ is executed before any other operation is executed on $\W$ or $\Z$.  Consequently, the durable objects $\W$ and $\Z$ behave like atomic EC objects, and are treated as such in the rest of the proof. 


We introduce the following terms. An {\em install} is a successful ECSC operation on $\W$, executed by a $\Write$ operation at its Line~14.
A $\Write$ operation {\em installs a value $v$ in $\W$ at time $t$} if it executes a successful ECSC on $\W$ at Line~14 at time $t$ and sets $\W.\val$ to $v$.

A {\em move} is a successful ECSC operation on $\Z$, executed by a $\transfer$ operation at its Line~20.
A $\transfer$ operation {\em moves a value $v$ to $\Z$ at time $t$} if it executes a successful ECSC on $\Z$ at Line~20 at time $t$ and sets $\Z.\val$ to $v$.

An {\em imprint} is a successful ECSC operation on $\Z$, executed by a $\CAS$ operation at its Line~10.
A $\CAS$ operation {\em imprints a value $v$ in $\Z$ at time $t$} if it executes a successful ECSC on $\Z$ at Line~14 at time $t$ and sets $\Z.\val$ to $v$.

A $\Write$ operation is {\em trivial} if it returns at Line~13*. 
A $\Write$ operation is a {\em hitchhiker} if it does not install and returns at Line~17.

To refer to the times when the algorithm performs various actions, we adopt the following notation.
If $\alpha$ is an execution of a $\Write$, $\CAS$, or $\transfer$ and $\ell$ is a line number in the algorithm, $\alpha[\ell]$ denotes the time at which $\alpha$ executes Line $\ell$.
Lines 7 through 10 may be executed twice because of the for-loop at Line~6*, and for any of these lines $\ell$, $\alpha[\ell,1]$ and $\alpha[\ell,2]$ denote the times at which a $\CAS$ operation $\alpha$ executes Line $\ell$ in the first and second iterations, respectively.
We denote an open interval from time $t_1$ to $t_2$ by $(t_1,t_2)$, the closed interval by $[t_1,t_2]$, and the interval that includes $t_1$ and not $t_2$ by $[t_1,t_2)$.

\begin{lemma}\label{inst1}
{\em
\mbox{ }

\begin{enumerate}
\item
If a $\Write{\h,\v}$ operation $\alpha$ installs, then
$\W$'s state remains the same in the interval $[\alpha[12], \alpha[14])$, and
at time $\alpha[14]$, $\W.\bit$ flips and $\W.\val$ becomes $v$.
\item
If a $\transfer{\h}$ operation $\alpha$ moves,
then $\Z$'s state remains the same in the interval $[\alpha[18], \alpha[20])$, and at time $\alpha[20]$, $\Z.\bit$ flips.
\item
If a $\CAS{\h, \old, \new}$ operation $\alpha$ imprints,
then $\Z$'s state remains the same in the interval $[\alpha[7,i], \alpha[10,i])$, 
where $i \in \{1,2\}$ is the iteration of the for-loop at Line~6* during which $\alpha$ imprints.
Furthermore, $\old \neq \new$ and, at time $\alpha[10,i]$, $\Z.\val$ changes from $\old$ to $\new$, and $\Z.\bit$ is unchanged.
\end{enumerate}
}
\end{lemma}

{\em Proof}: In Part (1), at time $\alpha[14]$, since the ECSC succeeds, it must be $\W.\seq = \w.\seq$. 
Since $\W.\seq$ was $\w.\seq$ back at $\alpha[12]$, it follows that $\W.\seq = \w.\seq$ all through the interval $[\alpha[12], \alpha[14])$, which implies that $\W$'s state remains the same all through that interval.
Since $\w = \W$ when the ECSC succeeds, the ECSC sets $\W.\val$ to $v$ and flips $\W.\bit$. Hence, we have Part (1).

Parts (2) and (3) are proved analogously.
Additionally, in Part (3), since $\alpha$ goes past Line~$[8,i]$ to execute the ECSC, we have $\z.\val = \old \neq \\new$. 
\hfill $\blacksquare$

The next lemma states that $\W$ changes and $\W.\bit$ flips precisely at times of installs, $\Z$ changes precisely at times of moves or imprints, and $\Z.\bit$ flips only at moves, and not at imprints.

\begin{lemma}\label{obs}
{\em
\mbox{ }
\begin{enumerate}
\item
At any time $t$, $\W$'s state changes or $\W.\bit$ flips at $t$ if and only if an install occurs at $t$.
\item
At any time $t$, $\Z$'s state changes at $t$ if only if a move or an imprint occurs at $t$.
\item
At any time $t$, $\Z.\bit$ flips at $t$ if and only if a move occurs at $t$.
\end{enumerate}
}
\end{lemma}

{\em Proof}: 
A successful ECSC on $\W$ at Line~14 changes $\W$'s state and flips $\W.\bit$, and no other action in the algorithm changes $\W$.
A successful ECSC on $\Z$ at Line~10 or Line~20 changes $\Z$'s state and no other action in the algorithm changes $\Z$; moreover, only a successful ECSC on $\Z$ at Line~20 flips $\Z.\bit$.
Hence, we have the lemma.
\hfill $\blacksquare$

\begin{lemma}\label{obs2}
{\em
\mbox{ }

\begin{enumerate}
\item
Suppose that $\alpha'$ and $\alpha$ are two different $\Write$ operations that install, and $\alpha'$ installs before $\alpha$, i.e., $\alpha'[14] < \alpha[14]$. Then, $\alpha'[14] < \alpha[12]$.
\item
Suppose that $\alpha$ and $\alpha'$ are two different $\transfer$ operations that move, and $\alpha'$ moves before $\alpha$, i.e., $\alpha'[20] < \alpha[20]$. Then, $\alpha'[20] < \alpha[18]$.
\item
Suppose that $\alpha'$ and $\alpha$ are two different $\CAS$ operations that imprint, and $\alpha'$ imprints at $\alpha'[10,i]$ before $\alpha$ imprints at $\alpha[10,j]$, for some $i, j \in \{1,2\}$. Then, $\alpha'[10,i] < \alpha[7,j]$.
\end{enumerate}
}
\end{lemma}

{\em Proof}: For a proof of Part (1), assume to the contrary that $\alpha[12] < \alpha'[14] < \alpha[14]$.
At $\alpha'[14]$, since the ECSC on $\W$ succeeds, $\W.\seq$ changes,
contradicting Part (1) of Lemma~\ref{inst1} that $\W$'s state remains the same in the interval $[\alpha[12], \alpha[14])$.
The other two parts of the lemma are proved analogously.
\hfill $\blacksquare$

\begin{lemma}\label{ifirst}
{\em
An install occurs before any move occurs.
}
\end{lemma}

{\em Proof}: Assume to the contrary that a move occurs at time $t$ and no install occurs before $t$.
Let $\alpha$ be the $\transfer$ operation that moves at $t$.
Since no installs or moves occur before $t$, Lemma~\ref{obs} implies that $\W.\bit$ and $\Z.\bit$ are never flipped before $t$.
So these bits still have their initial value of 0 when $\alpha$ executes ECLL at lines 18 and 19 to read the states of $\Z$ and $\W$ into $z$ and $w$, respectively.
Thus, $\z.\bit = \w.\bit = 0$ when $\alpha$ executes Line~20.
Therefore, the if-condition at Line~20 prevents $\alpha$ from executing the ECSC operation, contradicting that $\alpha$ moves.
\hfill $\blacksquare$

The next lemma states that installs and moves alternate.

\begin{lemma}\label{rhythm}
{\em
Between any two moves, there is an install; and between any two installs, there is a move.
}
\end{lemma}

{\em Proof}: We consider two cases---that installs occur consecutively with no intervening moves or that moves occur consecutively with no intervening installs---and derive a contradiction in each case.

Suppose that the lemma is violated for the first time when an install occurs at time $t$, following an install at an earlier time $t' < t$ and with no moves in the interval from $t'$ to $t$.
Since the lemma was not violated before $t$, Lemma~\ref{obs} implies that, counting $\W$'s flip at $t'$, $\W$ is flipped one more time than $\Z$ by time $t'$, and no further flips happen to either bit in the open interval from $t'$ to $t$.
It follows that, in the open interval from $t'$ to $t$, the values of $\W.\bit$ and $\Z.\bit$ do not change and, since $\W.\bit$ and $\Z.\bit$ are both initially 0, $\W.\bit \neq \Z.\bit$.
Let $b$ and $1-b$ be the stable values of $\W.\bit$ and $\Z.\bit$, respectively, in the interval $(t',t)$.
Let $\alpha'$ and $\alpha$ denote the $\Write$ operations that install at times $t'$ and $t$, respectively; thus, $t' = \alpha'[14]$ and $t = \alpha[14]$.
From Part (1) of Lemma~\ref{obs2}, we know that $\alpha'[14] < \alpha[12] < \alpha[14]$.
Then, since $\W.\bit$ and $\Z.\bit$ remain unchanged at $b$ and $1-b$ during the open interval $(\alpha'[14], \alpha[14])$, when $\alpha$ executes ECLL at Lines 12 and 13 to read $\W$'s state and $\Z$'s state into $w$ and $z$, respectively, $\w.\bit$ is set to $b$ and $\z.\bit$ is set to $1-b$.
Therefore, when $\alpha$ executes Line~14, the if-condition there evaluates to {\em false}, preventing $\alpha$ from executing the ECSC operation, which contradicts that $\alpha$ installs.
By an analogous argument, a contradiction arises in the second case also.
Hence, we have the lemma.
\hfill $\blacksquare$

\begin{corollary}\label{seesaw}
{\em
\mbox{ }

\begin{enumerate}
\item
$\W.\bit$ and $\Z.\bit$ do not change and have the same value in the open interval up to the first install and in each open interval from a move to the first install that follows that move.
\item
$\W.\bit$ and $\Z.\bit$ do not change and have different values in each open interval from an install to the first move that follows that install.
\end{enumerate}
}
\end{corollary}

{\em Proof}: Follows from Lemmas \ref{obs} and \ref{rhythm}, and the intialization of $\W.\bit$ and $\Z.\bit$ to 0.
\hfill $\blacksquare$

\begin{lemma}\label{snap}
{\em
\mbox{ }

\begin{enumerate}
\item
If a $\Write$ operation $\alpha$ installs, then $\W.\bit = \Z.\bit$ at time $\alpha[13]$.
\item
If a $\transfer$ operation $\alpha$ moves, then $\W.\bit \neq \Z.\bit$ at time $\alpha[19]$.
\end{enumerate}
}
\end{lemma}

{\em Proof}:
Since $\alpha$ is successful, it follows from Lemma~\ref{inst1} that $\W$'s state does not change in the open interval from $\alpha[12]$ to $\alpha[14]$.
Therefore, $\w$, which is $\W$'s state returned by $\alpha$'s ECLL at Line~12, continues to be $\W$'s state at $\alpha[14]$.
The ECLL on $\Z$ at Line~13 returns $\Z$'s state at $\alpha[10]$ into $\z$.
Thus, the values of $\W.\bit$ and $\Z.\bit$ at time $\alpha[10]$ are $\w.\bit$ and $\z.\bit$, respectively.
Since $\alpha$ installs (i.e., executes a successful ECSC at Line~14), the if-condition at Line~14 must evaluate to {\em true}, which implies that $\z.\bit = \w.\bit$, which from the above implies that $\W.\bit = \Z.\bit$ at time $\alpha[13]$.
Hence, we have Part (1).
Part (2) is proved analogously.
Hence, we have the lemma.
\hfill $\blacksquare$

\begin{lemma}\label{wtozhelp}
{\em
\mbox{ }

\begin{enumerate}
\item
If an install occurs at time $t$ and the first move after $t$ is executed by a $\transfer$ operation $\alpha$, then $\alpha[19] > t$.
\item
If a move occurs at time $t$, and the first install after $t$ is executed by a $\Write$ operation $\alpha$, then $\alpha[13] > t$.
\end{enumerate}
}
\end{lemma}

{\em Proof}: For a proof of Part (1), assume for a contradiction that $\alpha[19] < t < \alpha[20]$.
Part (2) of Lemma~\ref{snap} implies that $\W.\bit \neq \Z.\bit$ at time $\alpha[19]$.
Therefore, the instant $\alpha[19]$ falls in a time interval where $\W.\bit \neq \Z.\bit$.
If the install at $t$ is the first install, let $t' = 0$; otherwise, let $t'$ be the latest time before $t$ when a move occurs.
Corollary~\ref{seesaw} implies that $\W.\bit = \Z.\bit$ during the open interval from $t'$ to $t$, and $\W.\bit \neq \Z.\bit$ during the open interval from $t$ to $\alpha[20]$.
Then, since $\alpha[19] < t$ and $\alpha[19]$ falls in a time interval where $\W.\bit \neq \Z.\bit$,
it must be that $\alpha[19] < t'$.
It follows that $t' > 0$ and the install at $t$ is not the first install.
From the above, we have $\alpha[19] < t' < t < \alpha[20]$.
Since a move occurs at $t'$, there is a successful ECSC on $\Z$ at $t'$.
Therefore, $\Z$'s state changes at $t'$, which implies that
$\Z$'s state changes between $\alpha[19]$ and $\alpha[20]$, contradicting Part (2) of Lemma~\ref{inst1}, which states that $\Z$'s state does not change in the interval from $\alpha[18]$ to $\alpha[20]$.
Hence, we have Part (1).
Part (2) is proved analogously.
Hence, we have the lemma.
\hfill $\blacksquare$

The next lemma states that if two $\transfer$ operations are executed after installing a value $v$ in $\W$, $v$ is sure to move to $\Z$ by the time the second $\transfer$ operation completes.

\begin{lemma}\label{tran}
{\em
If $\W.\bit \neq \Z.\bit$ at time $t$,
a $\transfer$ operation $\alpha_1$ is started after $t$, and another $\transfer$ operation $\alpha_2$ is started after $\alpha_1$ completes, then a move occurs  after $t$ and at or before $\alpha_2$'s completion time of $\alpha_2[20]$.
}
\end{lemma}

{\em Proof}: Assume to the contrary that no move occurs between $t$ and $\alpha_2[20]$.
Then, Corollary~\ref{seesaw} implies that, in the interval from $t$ to $\alpha_2[20]$, $\W.\bit$ and $\Z.\bit$ do not change and $\W.\bit \neq \Z.\bit$.
Let $b \in \{0,1\}$ and $1-b$ be $\W.\bit$'s and $\Z.\bit$'s values, respectively, in this interval.
Therefore, in both $\alpha_1$ and $\alpha_2$, the ECLL operations on $\Z$ and $\W$ at Lines 18 and 19 return $z$ and $w$ such that $z.\bit = 1-b$ and $w.\bit = b$;
so, the if-condition at Line~20 evaluates to {\em true} and the ECSC operation on $\Z$ is attempted.
However, by our assumption that no move occurs by time $\alpha_2[20]$, the ECSC on $\Z$ at Line~20 is unsuccessful in both $\alpha_1$ and $\alpha_2$.

The failure of $\alpha_1$'s ECSC at Line~20 must be because $\Z$'s state changed between $\alpha_1$'s ECLL at Line~18 and its subsequent ECSC at Line~20.
Therefore, by Lemma~\ref{obs}, an imprint or a move occurs between $\alpha_1[18]$ and $\alpha_1[20]$.
It must be an imprint because a move is ruled out by our assumption that no move occurs by time $\alpha_2[20]$.
Let $\beta_1$ be the $\CAS$ operation that imprints between $\alpha_1[18]$ and $\alpha_1[20]$.
Arguing similarly, there is a $\CAS$ operation $\beta_2$ that imprints between $\alpha_2[18]$ and $\alpha_2[20]$.
Let $i_1 \in \{1,2\}$ and $i_2 \in \{1,2\}$ be the iterations (of the for-loop at Line~6*) during which $\beta_1$ and $\beta_2$, respectively, execute a successful ECSC on $\Z$.
Part (3) of Lemma~\ref{inst1} implies that $\Z$'s state is unchanged in the open interval from $\beta_2[7,i_2]$ to $\beta_2[10,i_2]$.

Since $\beta_1$ imprints between $\alpha_1[18]$ and $\alpha_1[20]$, we have $\alpha_1[18] < \beta_1[10,i_1] < \alpha_1[20]$.
Similarly, we have $\alpha_2[18] < \beta_2[10,i_2] < \alpha_2[20]$.
Furthermore, Part (3) of Lemma~\ref{obs2} implies that $\beta_1[10,i_1] < \beta_2[7,i_2]$.
Putting these together, we have $t < \alpha_1[18] < \beta_1[10,i_1] < \beta_2[7,i_2] < \beta_2[10,i_2] < \alpha_2[20]$.
In particular, $t < \beta_2[7,i_2] < \beta_2[10,i_2] < \alpha_2[20]$.
Furthermore, Part (3) of Lemma~\ref{inst1} implies that $\Z$'s state is unchanged in the open interval from $\beta_2[7,i_2]$ to $\beta_2[10,i_2]$; in particular, $\Z$'s state is constant throughout the time that $\beta_2$ calls the $\transfer$ method from its Line$[9,i_2]$ and executes it.
The last two observations imply that, during this execution of $\transfer$,
$\beta_2$'s ECLL on $\Z$ at Line~18 returns a $\zh$ such that $\zh.\bit = 1-b$, its ECLL on $\W$ at Line~19 returns a $\wh$ such that $\wh.\bit = b$, and $\Z$'s state is still $\zh$ when $\beta_2$ executes Line~20.
It follows that the if-condition at Line~20 evaluates to {\em true}, so $\beta_2$ executes the ECSC operation; and this ECSC succeeds because $\Z$'s state is still $\zh$.
This successful ECSC means that $\beta_2$ moves, contradicting our assumption that no move occurs between $t$ and $\alpha_2[20]$.
\hfill $\blacksquare$

The next lemma states that if a $\Write$ operation $\alpha$ installs a value $v$ in $\W$, that value is moved to $\Z$ before $\alpha$ completes.

\begin{lemma}\label{wok}
{\em
If a $\Write{\h,\v}$ operation $\alpha$ installs $\v$ at time $t$, the first move after $t$ occurs before $\alpha$ completes and it moves $\v$ to $\Z$.
}
\end{lemma}

{\em Proof}: After $\alpha$ installs at time $t$ (by executing a successful ECSC on $\W$ at Line~14), $\alpha$ executes the $\transfer$ method twice, at Lines 15 and 16.
If the process $p$ executing $\alpha$ crashes before executing the $\transfer$ method twice, upon restart, $p$ executes the {\sc recover} method where, after recovering from any partially executed operations on $\W$ and $\Z$ at Lines 21 to 24, $\transfer$ is executed twice, at Lines 25 and 26.
Thus, after $\alpha$ installs at time $t$, regardless of crashes, $\alpha$ executes $\transfer$ at least twice, before completing.
It follows from Lemma~\ref{tran} that a move occurs before $\alpha$ is completed.
Hence, we have the first part of the lemma.

To complete the lemma, we prove that the first move after $t$ moves $\v$ to $\Z$.
Let $\alpha'$ be the $\transfer$ operation that executes the first move after $t$.
$\W.\val$ is set to $v$ at time $t$ (by Part (1) of Lemma~\ref{inst1}), and it remains unchanged until the next install (by Part (1) of Lemma~\ref{obs}).
Since a move must occur between any two installs (by Lemma~\ref{rhythm}), it follows that $\W.\val = \v$ during the interval from $t$ to $\alpha'[20]$.
Furthermore, $\alpha'[19] > t$ (by Lemma~\ref{wtozhelp}).
It follows from the above that, when $\alpha'$ reads $\W$ into $\wh$ at Line~19, $\wh.\val = \v$.
Therefore, when $\alpha'$ moves by executing a successful ECSC on $\Z$ at Line~20, the passing of $\wh.\val$ as the third argument of that ECSC ensures that $\Z.\val$ is set to $\v$.
Thus, the first move after $t$ moves $v$ to $\Z$.
Hence, we have the lemma.
\hfill $\blacksquare$

The next lemma states that if a value $v$ is moved to $\Z$ and there are no imprints after the move, then the next installed value can't be $v$.
This property is crucial to ensuring the correctness and wait-freedom of CAS operations in the face of an unbounded stream of write operations that write the same value.

\begin{lemma}\label{norepeat}
{\em
If a move occurs at time $t'$, an install occurs at time $t > t'$, and no imprints or moves occur between $t'$ and $t$, then the value installed at $t$ differs from the value moved at $t'$.
}
\end{lemma}

{\em Proof}: Let $v'$ be the value moved at $t'$ and $\alpha$ be the $\Write{\h,\v}$ operation that installs at $t$.
Assume that no imprints or moves occur between $t'$ and $t$. 
We need to prove that $v \neq v'$.
The move at $t'$ sets $\Z.\val$ to $v'$, and
since no moves or imprints occur in the open interval from $t'$ to $t$, where $t = \alpha[14]$, 
$\Z.\val$ remains at $v'$ throughout this interval.
Furthermore, $\alpha[13] > t'$ (by Part (2) of Lemma~\ref{wtozhelp}).
Therefore, $\alpha$'s ECLL on $\Z$ at Line~13 returns $\z$ such that $\z.\val = \v'$.
Therefore, if $v'$ were the same as $v$, the if-condition at Line~13* would cause $\alpha$ to return at Line~13*, contradicting that $\alpha$ installs.
We conclude that $v' \neq v$.
\hfill $\blacksquare$

The next lemma assures that a CAS operation, which returns $\false$ at Line~11, can be linearized.
Its proof brings out the purpose of the two distinctive features of DuraCAS not shared by DurECW---Line~13* (using which the previous lemma was proved) and iterating twice through lines 7 to 10.

\begin{lemma}\label{twoit}
{\em
If a $\CAS{\h, \old, \new}$ operation $\alpha$ returns $\false$ at Line~11, there exists a time $t$ in the interval $I = (\alpha[7,1], \alpha[11])$ during which $\alpha$ executes such that $\Z.\val \neq \old$ at $t$.
}
\end{lemma}

\noindent{\em Proof}: We consider three cases.

\begin{itemize}
\item
\underline{Case 1}:
Suppose that some $\CAS{\h', u, v}$ operation imprints at some time $\tau$ during the interval $I$.
It follows from Part (3) of Lemma~\ref{inst1} that $u \neq v$, and $\Z.\val$ changes from $u$ to $v$ at $\tau$.
Since $u \neq v$, one of $u$ and $v$ must be different from $\old$.
Hence, the lemma holds in Case 1.
\item
\underline{Case 2}:
Suppose that there no imprints during $I$, but at least two moves occur during the interval $I$.
Let $u$ be the value moved by the first move $M_1$ that occurs during $I$, and $v$ be the value installed by the first install after the move $M_1$.
Then, since moves and installs alternate (Lemma~\ref{rhythm}), the move $M_2$ that occurs after this install would be the second move to occur in the interval $I$, and it moves $v$ (Lemma~\ref{wok}). 
Lemma~\ref{norepeat} assures that the values $u$ and $v$, moved by $M_1$ and $M_2$, are different.
Therefore, one of $u$ and $v$ must be different from $\old$.
Hence, the lemma holds in Case 2.
\item
\underline{Case 3}:
Suppose that neither of the above two cases holds, i.e., during the interval $I$, no imprints occur and at most one move occurs.
Then, no move occurs during at least one of the two iterations of Lines 7 to 10.
Let $i \in \{1,2\}$ be the iteration of the for-loop at Line~6* during which no move occurs.
Since there are also no imprints during this iteration, $\Z$'s state at Line~$[10,i]$ is the same as $\Z$'s state read into $\z$ at Line~$[7,i]$.
Therefore, the ECSC at Line~10 succeeds, causing $\alpha$ to return $\true$ at Line~10*, which contradicts that $\alpha$ returns $\false$.
Hence, Case 3 does not arise. 
\end{itemize}

Hence, we have the lemma.
\hfill $\blacksquare$

Recall that a hitchhiker is a $\Write$ operation that does not install and returns at Line~17.

\begin{lemma}\label{hh1}
{\em
If $\alpha$ is a hitchhiker $\Write$ operation, $\W.\bit \neq \Z.\bit$ at some time during the semi-closed interval $(\alpha[12], \alpha[14]]$.
}
\end{lemma}

{\em Proof}: Assume to the contrary that $\W.\bit = \Z.\bit$ throughout the interval $(\alpha[12], \alpha[14]]$.
Then, Corollary~\ref{seesaw} implies that in this interval $\W$'s state and $\Z.\bit$ do not change their values.
It follows that, when $\alpha$ executes an ECLL on $\W$ and $\Z$ at Lines 12 and 13, the return values $\w$ and $\z$ are such that $\w.\bit = \z.\bit$ and $\W$'s state continues to be $\w$ at $\alpha$'s Line~14.
Since $\alpha$ is a hitchhiker, it does not return at Line~13*;
furthermore, at Line~14, since $\w.\bit = \z.\bit$, the if-condition evaluates to {\em true}, and since $\W = \w$, the ECSC succeeds.
Thus, $\alpha$ installs, contradicting that $\alpha$ is a hitchhiker.
\hfill $\blacksquare$

The next lemma states that a move occurs during every hitchhiking write.

\begin{lemma}\label{hh}
{\em
If $\alpha$ is a hitchhiker $\Write$ operation, a move occurs between $\alpha[12]$ and $\alpha[17]$.
}
\end{lemma}

{\em Proof}: By the previous lemma, $\W.\bit \neq \Z.\bit$ at some time $t$ between $\alpha[12]$ and $\alpha[14]$. Since $\alpha$ executes two $\transfer$ operations after $\alpha[14]$ and before it returns at Line~17, Lemma~\ref{tran} implies that a move occurs between $t$ and $\alpha[17]$.
Hence, we have the lemma.
\hfill $\blacksquare$

The next definition states how a $\Write{\h,\v}$ operation is linearized, based on whether it installs, hitchhikes, or is trivial.
(Recall that a a $\Write{\h,\v}$ operation is trivial if it returns at Line~13*.)
A crashed operation is not linearized, unless it is a $\Write$ operation that crashes after installing.  
Note that trivial operations return at Line~13* and hitchhikers return at Line~17, so they are not crashed operations.

\begin{definition}[Linearization of Write]\label{wlin}
{\em
Let $\alpha$ be a $\Write{\h,\v}$ operation.

\begin{enumerate}
\item
If $\alpha$ is trivial, it is linearized at $\alpha[13]$, the time when $\alpha$ executes Line~13.
\item
If $\alpha$ installs, it is linearized at the first move after $\alpha$'s install.

(Lemma~\ref{wok} guarantees that $\alpha$ is linearized before it completes.)

\item
If $\alpha$ is a hitchhiker, let $t$ be the earliest time when a move occurs in $\alpha$'s interval, and $\beta$ be the unique installer linearized at $t$.
(Lemma~\ref{hh} guarantees that $t$ is well defined, and Lemma~\ref{rhythm} guarantees that a unique installer is linearized at $t$.)
Then, $\alpha$ is linearized at $t$ (along with $\beta$), and is ordered {\em before} $\beta$.
(This ordering ensures that the hitchhikers are overwritten instantly by the installer, thereby eliminating the burden of detecting the hitchhikers' write operations.)
\end{enumerate}
}
\hfill $\blacksquare$
\end{definition}

Next we state how $\CAS$ and $\Read$ operations are linearized.
We choose not to linearize a crashed $\CAS$ operation, unless it crashes after imprinting.

\begin{definition}[Linearization of CAS and Read]\label{clin}
{\em
\mbox{ }

\begin{enumerate}
\item
If a $\CAS{\h, \old, \new}$ operation imprints, it is linearized at the time it imprints.
\item
If a $\CAS{\h, \old, \new}$ operation returns $\false$ at Line~11, it is linearized at the earliest time when $\Z.\val \neq \old$.

(Lemma~\ref{twoit} guarantees that $\alpha$ is linearized during the time interval in which it executes.)

\item
If a $\CAS{\h, \old, \new}$ operation $\alpha$ returns at Line~$[8,i]$, $i \in \{1,2\}$, it is linearized at $\alpha[7,i]$.
\item
If a $\Read{\h}$ operation $\alpha$ returns, it is linearized at $\alpha[4]$.
\end{enumerate}
}
\hfill $\blacksquare$
\end{definition}

The value of a DuraCAS object implemented by the algorithm changes atomically at points where $\Write$ and $\CAS$ operations are linearized.
The next lemma states that the algorithm maintains the DuraCAS object's value in $\Z.\val$, and satisfies durable linearizability.

\begin{lemma}[Durable-linearizability of \DuraCAS \ objects]\label{juice}
{\em 
Let $\O$ be a DuraCAS object implemented by the algorithm.

\begin{enumerate}
\item
$\O.\val = \Z.\val$ at all times.
\item
Let $\alpha$ be any $\CAS{\h, \old,\new}$, $\Write{\h, \v}$, or $\Read{\h}$ operation, and $t$ be the time at which $\alpha$ is linearized.
Suppose that $\O.\val = \sigma$ at $t$ and just before $\alpha$'s linearization (in case multiple operations are linearized at $t$), and $\delta(\sigma, \alpha) = (\sigma', r)$, where $\delta$ is the sequential specification of a CAS object. Then:

\begin{enumerate}
\item
$\O.\val$ changes to $\sigma'$ at time $t$.
\item
If $\alpha$ completes without crashing, it returns $r$.

(Recall that if $\alpha$ crashes and, upon restart, executes $\Recover$, the recover method does not return any response.)
\end{enumerate}

\end{enumerate}
}
\end{lemma}

{\em Proof}: 
We prove the lemma by induction.
The base step follows from the algorithm's initialization that sets $\Z.\val$ to $\mathcal O$'s initial value.
The induction hypothesis is that $i \ge 1$ and the lemma holds up to and including the first $i-1$ linearization times.
Let $\tau$ be the next linearization time and $S$ be the set of operations that are linearized at $\tau$.
For the induction step, we show that the lemma holds even after the operations in $S$ take effect at $\tau$ (in their linearization order).
There are many possibilities for what operations are linearized at $\tau$, and we show the induction step for each possibility.

\begin{itemize}
\item
\underline{Case 1, a move occurs at $\tau$}:
In this case, a set $S$ of hitchhiker $\Write$ operations and one installing $\Write{\h,\v}$ operation $\alpha$ are linearized, all at $\tau$, with $\alpha$ linearizing after the ones in $S$ (by Definition~\ref{wlin}). As a result, all hitchhiker writes are overwritten by $\alpha$, and $\O.\val$ becomes $v$ after all of the operations are linearized. 
Furthermore, the move at $\tau$ sets $\Z.\val$ to $v$ (by Lemma~\ref{wok}), thereby ensuring that $\O.\val = \Z.\val$ immediately after $\tau$.


\item
\underline{Case 2, an imprint occurs at $\tau$}:
In this case, the $\CAS{\h, \old, \new}$ operation $\alpha$ that imprints at $\tau$ is the only operation linearized at $\tau$.
Then, $\Z.\val$ changes from $\old$ to $\new$ at $\tau$ (by Part (3) of Lemma~\ref{inst1}).
By the induction hypothesis, immediately before $\alpha$'s linearization at $\tau$, $\O.\val = \Z.\val = old$, and $\alpha$'s linearization changes $\O.\val$ from $\old$ to $\new$.
Furthermore, if $\alpha$ does not crash after imprinting (i.e., after the successful ECSC at Line~10), it returns the correct response of {\em true} at Line~10*.

\item
\underline{Case 3, a $\CAS{\h, \old, \new}$ operation $\alpha$, which returns $\false$ at Line~11, is linearized at $\tau$}:
In this case, Part (2) of Definition~\ref{clin} implies that $\Z.\val \neq \old$ at $\tau$.
By the induction hypothesis, $\O.\val \neq \old$ at $\tau$.
Therefore, $\alpha$'s linearization at $\tau$ does not change $\O.\val$ at $\tau$, and $\O$ returns $\false$ to $\alpha$.
Since $\Z$ is not changed at $\tau$ and $\alpha$ returns $\false$ at Line~11, the induction step holds for this case.

\item
\underline{Case 4, a trivial write is linearized at $\tau$}:
In this case, a $\Write{\h,\v}$ operation $\alpha$, which reads $\v$ in $\Z.\val$  at Line~13 at time $\tau$ and returns at the next line, is linearized at $\tau$.
Thus, immediately before $\alpha$'s linearization at $\tau$, $\Z.\val = v$ and, by the induction hypothesis, $\O.\val = v$. 
Immediately after $\alpha$'s linearization at $\tau$, $\Z.\val$ continues to be $v$ (since $\Z$ is not changed at Line~13) and $\O.\val$ also continues to be $\v$ because $\alpha$'s $\Write{\h,\v}$ operation takes effect at $\tau$, changing $\O.\val$ from $\v$ to $\v$.

\item
\underline{Case 5, a $\CAS{\h, \old, \new}$ operation $\alpha$, which returns at Line~8, is linearized at $\tau$}:
In this case, by Definition~\ref{clin}, $\tau$ is the time of $\alpha$'s execution of Line~7.
It follows from the code at Lines 7 and 8 that at $\tau$, $\Z.\val$ holds some $v$ such that either $\v \neq \old$ or $\v = \old = \new$.
By the induction hypothesis, $\O.\val = \v$ at $\tau$, so $\alpha$'s linearization at $\tau$ implies that $\O.\val$ remains unchanged, and ${\O}$'s response is $\false$ if $\v \neq \old$, and $\true$ otherwise. 
This justifies $\alpha$'s return at Line~8, without changing $\Z$. 

\item
\underline{Case 6, a $\Read{\h}$ operation $\alpha$ is linearized at $\tau$}:
In this case, $\tau$ is the time of $\alpha$'s execution of Line~4.
Since $\O.\val = \Z.\val$ at $\tau$ (by the induction hypothesis), $\alpha$'s linearization at $\tau$ implies that ${\mathcal O}.\val$ remains unchanged, and ${\mathcal O}$'s response is $\Z.\val$ at $\tau$, which is $z.\val$.
This justifies $\alpha$ returning $z.\val$ at Line~5, without changing $\Z$.
\end{itemize}

Hence, the induction step is complete and we have the lemma.
\hfill $\blacksquare$

Next we prove that a DuraCAS object $\O$ implemented by the algorithm is detectable.
The key to achieving detectability lies in limiting the use of the handle $\h.\Critical$ to Lines 14 and 10, where install and imprint are attempted.
In particular, when a move is attempted at Line~20, the algorithm employs $\h.\Casual$, and not $\h.\Critical$.
This discrimination ensures that a visible operation---an installing write or an imprinting CAS that affect $\O$'s state in a manner that future operations might witness---increases the ``detector value'' associated with the handle $\h.\Critical$, while safe-to-repeat operations---hitchhiking and trivial writes, reads, CAS operations that return without affecting $\O$'s state (like the ones that return at Line~8 or return $\false$ at Line~10), or operations that crash without affecting $\O$'s state---do not increase the detector value associated with $\h.\Critical$.

\begin{lemma}[Detectability of \DuraCAS \ objects]\label{det}
{\em
Let $\alpha$ be any operation executed on a DuraCAS object $\O$ by a handle $\h$.
Suppose that $(d_1, r_1)$ and $(d_2, r_2)$ are the values that $\Detect{\h}$ would return, if executed immediately before $\alpha$ is invoked and immediately after $\alpha$ completes, respectively.
Then:
\begin{enumerate}
    \item
    If $\alpha$ is neither an installing write nor an imprinting CAS, it is safe to repeat and $d_2 = d_1$.
    \item
    If $\alpha$ is an installing write or an imprinting CAS, then $d_2 > d_1$ and $r_2 = \true$.
\end{enumerate}
}
\end{lemma}

{\em Proof}:  
For Part (1), a hitchhiking write operation changes $\O$'s state, but the change is rendered invisible to all future operations because it is instantly overwritten by an installing writer.
Trivial writes, reads, CAS operations that return at Line~8, and CAS operations that return $\false$ at Line~10 do not change $\O$'s state.
Hence, all of these operations are safe to repeat.
Furthermore, none of these operations perform a successful ECSC on $\W$ or $\Z$ using the handle $\h.\Critical$
(they might perform a successful ECSC at Line~20, but the handle used there is not $\h.\Critical$).
Therefore, DurEC's detectability property implies that calls to $\Detect{\h.\Critical}$ before and after $\alpha$ would return $(d_1, -)$ and $(d_2, -)$ such that $d_1 = d_2$.
So, the call to $\Detect{\h.\Critical}$ at Line~27 establishes the second half of Part (1) of the lemma.

For Part (2), suppose that $\alpha$ is an installing write or an imprinting CAS.
In both cases, $\alpha$ performs a successful ECSC operation $op$ on one of $\W$ or $\Z$, using the handle $\h.\Critical$ (this happens at Line~14 if $\alpha$ is an installing write, and at Line~10 if it is an imprinting CAS).
Then, by the detectability of DurEC objects $\W$ and $\Z$, it follows that 
$\Detect{\h.\Critical}$, if executed before $op$ and after $op$ would return $d_1$ and $d_2$ such that $d_2 > d_1$.
This fact, together with how $\Detect{\h}$ is implemented by Line~27, imply Part (2) of the lemma.
\hfill $\blacksquare$

\durlin*

{\em Proof}: The first property was proved in Lemmas \ref{juice} and \ref{det}.
The second property is clear from an inspection of the algorithm.
The third property follows from the observation that the algorithm makes no assumption about the the number or the names of processes that participate in the algorithm: any process can begin participating by creating a handle for itself, and accessing any existing DuraCAS objects, or creating new ones by calling the constructor.
For the fourth property, we note that
each DuraCAS handle $\h$ needs space for two DurEC handles ($\h.\Critical$ and $\h.\Casual$), and
each DuraCAS object $\O$ needs space for two DurEC objects ($\W$ and $\Z$).
Since each DurEC handle and DurEC object requires only $O(1)$ space, if $n$ DuraCAS handles and $m$ DuraCAS objects are created in a run, the space required is $O(m+n)$.
\hfill $\blacksquare$



\end{document}